\documentclass{aa}
\usepackage[usenames,dvipsnames]{xcolor}

\usepackage{graphicx}
\usepackage{amsmath, bm}
\usepackage{natbib}
\bibpunct{(}{)}{;}{a}{}{,} 

\usepackage{tikz}
\usepackage{txfonts}
\usepackage{longtable}
\usepackage{listings}
\usepackage{color}
\usepackage{textcomp}
\usepackage{multirow}

\usepackage{soul}
\usepackage[normalem]{ulem}  

\begin{document}

\title{Spectroscopic age estimates for APOGEE red-giant stars\thanks{Table \ref{datamodel} is only available in electronic form
at the CDS via 
\texttt{https://cdsarc.cds.unistra.fr/cgi-bin/qcat?J/A+A/}.}: \\ Precise spatial and kinematic trends with age in the Galactic disc
}

\author{F. Anders\inst{1,2,3}, 
        P. Gispert\inst{1}, 
        B. Ratcliffe\inst{4}, C. Chiappini\inst{4}, I. Minchev\inst{4}, \\ S. Nepal\inst{4,5}, A. B. A. Queiroz\inst{4,5}, J. A. S. Amarante\inst{1,2,3,6}, T. Antoja\inst{1,2,3}, G. Casali\inst{7,8}, \\ L. Casamiquela\inst{1,2,3,9}, A. Khalatyan\inst{4}, A. Miglio\inst{5,6}, H. Perottoni\inst{1,10}, M. Schultheis\inst{12}
        }
    
\institute{Departament de Física Quàntica i Astrofísica (FQA), Universitat de Barcelona (UB),  Martí i Franquès, 1, 08028 Barcelona, Spain
    \and{Institut de Ciències del Cosmos (ICCUB), Universitat de Barcelona (UB), Martí i Franquès, 1, 08028 Barcelona, Spain}
    \and{Institut d'Estudis Espacials de Catalunya (IEEC), Gran Capità, 2-4, 08034 Barcelona, Spain}
            \email{fanders@fqa.ub.edu}
    \and{Leibniz-Institut für Astrophysik Potsdam (AIP), An der Sternwarte 16, 14482 Potsdam, Germany}
    \and{Institut für Physik und Astronomie, Universität Potsdam, Haus 28 Karl-Liebknecht-Str. 24/25, D-14476 Golm, Germany}
    \and{Visiting Fellow, Jeremiah Horrocks Institute, University of Central Lancashire, Preston, PR1 2HE, UK}
    \and{Department of Physics \& Astronomy "Augusto Righi", University of Bologna, via Gobetti 93/2, 40129 Bologna, Italy}
    \and{INAF-Astrophysics and Space Science Observatory of Bologna, via Gobetti 93/3, 40129 Bologna, Italy}
    \and{GEPI, Observatoire de Paris, PSL Research University, CNRS, Sorbonne Paris Cité, 5 place Jules Janssen, 92190 Meudon, France}
    \and{Universidade de S\~{a}o Paulo, Instituto de Astronomia, Geof\'isica e Ci\^{e}ncias Atmosf\'ericas, Departamento de Astronomia, SP 05508-090, S\~{a}o Paulo, Brasil}
    \and{Université Côte d’Azur, Observatoire de la Côte d’Azur, Laboratoire Lagrange, CNRS, Blvd de l’Observatoire, F-06304 Nice,
France}
           }

\date{Received \today; accepted ...}

  \abstract{Over the last few years, many studies have found an empirical relationship between the abundance of a star and its age. Here we estimate spectroscopic stellar ages for 178\,825 red-giant stars observed by the APOGEE survey with a median statistical uncertainty of 17$\%$. To this end, we use the supervised machine learning technique {\tt XGBoost}, trained on a high-quality dataset of 3\,060 red-giant and red-clump stars with asteroseismic ages observed by both APOGEE and {\it Kepler}. 
  After verifying the obtained age estimates with independent catalogues, we investigate some of the classical chemical, positional, and kinematic relationships of the stars as a function of their age. We find a very clear imprint of the outer-disc flare in the age maps and confirm the recently found split in the local age-metallicity relation. We present new and precise measurements of the Galactic radial metallicity gradient in small age bins between 0.5 and 12 Gyr, confirming a steeper metallicity gradient for $\sim2-5$ Gyr old populations and a subsequent flattening for older populations mostly produced by radial migration. In addition, we analyse the dispersion about the abundance gradient as a function of age. We find a clear power-law trend (with an exponent $\beta\approx0.15$) for this relation, indicating a relatively smooth radial migration history in the Galactic disc over the past $7-9$ Gyr. Departures from this power law may possibly be related to the Gaia Enceladus merger and passages of the Sagittarius dSph galaxy. 
  Finally, we confirm previous measurements showing a steepening in the age-velocity dispersion relation at around $\sim9$ Gyr, but now extending it over a large extent of the Galactic disc (5 kpc $<R_{\rm Gal}<13$ kpc). To establish whether this steepening is the imprint of a Galactic merger event, however, detailed forward modelling work of our data is necessary. Our catalogue of precise stellar ages and the source code to create it are publicly available.
  }

\titlerunning{Spectroscopic ages of red-giant stars}
\authorrunning{Anders, Gispert, Ratcliffe, et al.}
\maketitle

\section{Introduction}
\label{sec:intro}

Isochrone fitting is frequently used to determine stellar ages of (primarily FGK) field stars, especially in the context of large spectroscopic stellar surveys (e.g. \citealt{Santiago2016, Mints2017, McMillan2018, Sanders2018, Lebreton2020, Queiroz2023}. While this technique has a long tradition (e.g. \citealt{Pont2004, Jorgensen2005}) and works reasonably well for stars close to the main-sequence turn-off and sub-giant branch (but see \citealt{Valle2013, Lebreton2014}), isochrone-fitting for red-giant stars is much more challenging and prone to sizeable statistical and systematic errors \citep{Soderblom2010, Noels2015}.

Another well-tested (but also model-dependent) method to estimate ages for field stars, including red giants, is offered by asteroseismology \citep{Chaplin2013}. Red-giant stars are particularly interesting for Galactic archaeology studies, since they are numerous, bright, and cover a wide range of stellar ages \citep{Miglio2012}. Until the upcoming PLATO mission (\citealt{Rauer2014}, \citealt{Miglio2017}), however, red-giant ages derived from joint asteroseismic and spectroscopic constraints are only available for select samples ($\lesssim10\,000$ stars) in certain fields, such as the {\it Kepler} (e.g. \citealt{Pinsonneault2014, Pinsonneault2018, Wu2018, Miglio2021, Matsuno2021}), CoRoT \citep{Valentini2016, Anders2017}, K2 (e.g. \citealt{Rendle2019, Valentini2019, Zinn2020, Zinn2022, Schonhut-Stasik2023}), or TESS Continuous Viewing Zone \citep{Sharma2018, SilvaAguirre2020, Mackereth2021, Wu2023} fields. 
Therefore, large-scale spectroscopic surveys such as APOGEE \citep{Majewski2017}, GALAH \citep{DeSilva2015}, LAMOST \citep{Cui2012}, or {\it Gaia} RVS \citep{GaiaCollaboration2022Recio, Recio-Blanco2022}, have been aiming at providing empirical spectroscopy-based age estimates for Galactic archaeology studies, typically using asteroseismology as benchmark data \citep[e.g.][]{Martig2016, Leung2019, Ciuca2022, He2022}. 

It has long been suggested that, in the absence of reliable stellar ages, precise chemical abundances could be used to determine the approximate age of a star. The works of \citet{Nissen2015, Nissen2016} and \citet{TucciMaia2016} showed that, at least for solar twins observed at high spectral resolution and high signal-to-noise ratio, this is a viable assumption. The tight relation between [Y/Mg] and stellar age found in these works, later confirmed by \citet{Spina2018}, demonstrated that a combination of elemental abundances probing different nucleosynthetic channels (in the case of [Y/Mg] an s-process and an $\alpha$ element) may provide a meaningful empirical measure for stellar age. Follow-up works also explored other elemental abundance ratios (e.g. \citealt{Jofre2020, Casamiquela2021}) and showed that when leaving the regime of stellar twins, this empirical "chemical clock" needs to include other terms (such as $T_{\rm eff}$ or [Fe/H]) to provide meaningful results (e.g. \citealt{Feltzing2017, DelgadoMena2019, Casali2020, Viscasillas2022}).

The APOGEE survey, which has taken high-resolution ($R\simeq22\,500$) high signal-to-noise ($\gtrsim70$) spectra of mainly red giant stars in the near-infrared $H$-band \citep{Wilson2019}, covering several carbon and nitrogen molecular bands \citep{AllendePrieto2008}, triggered the discovery that in red-giant stars the [C/N] ratio correlates with stellar mass (and therefore age; \citealt{Masseron2015}). Subsequent theoretical work by \citet{Salaris2015} showed that [C/N] measurements alone cannot provide precise and accurate ages for individual stars, but may well provide population ages (see also \citealt{Lagarde2017} for a detailed investigation of the underlying mixing effects that produce the [C/N] dependence on age). The [C/N] ratio was since used with success to estimate red-giant ages (with $\simeq30-40\%$ precision) through a calibration of the [C/N]-age relation with asteroseismic data from the {\it Kepler} mission \citep{Martig2016, Ness2016, Wu2019, Huang2020, Zhang2021} or with open clusters \citep{Casali2019, Spoo2022}. 

\begin{figure}
\begin{center} 
\includegraphics[width=.495\textwidth]{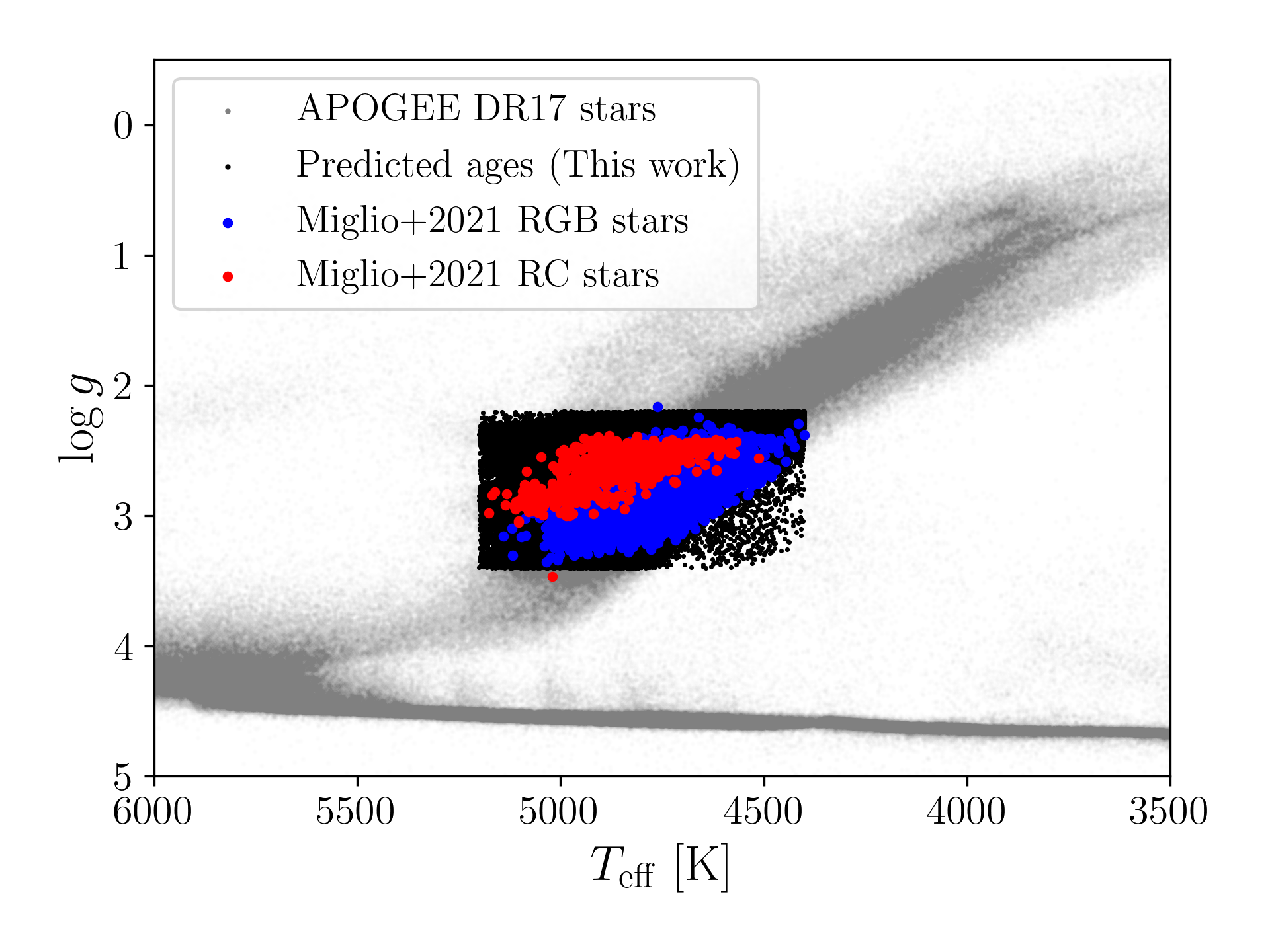}\\
\includegraphics[width=.495\textwidth]{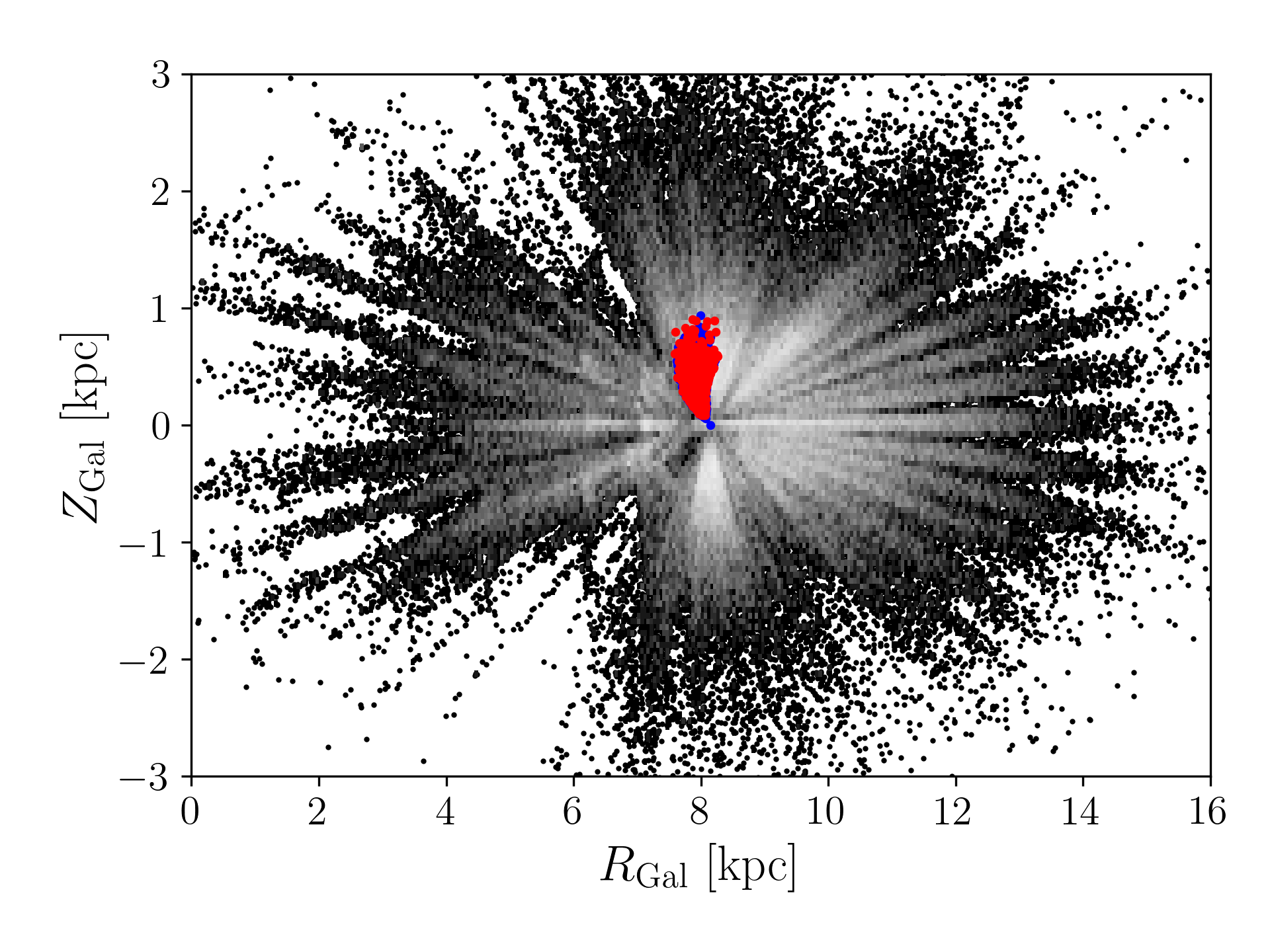}\\
\caption{\label{fig:kiel} 
Parameter space covered by our APOGEE sample. Top panel: Kiel diagram ($\log g$ vs. $T_{\rm eff}$) for the APOGEE DR17
catalogue (grey density). The black dots highlight the stars range for which our method provides chemical stellar age estimates.
Bottom panel: Galactic distribution of the 178\,825 APOGEE stars with spectroscopic ages in cylindrical coordinates ($Z_{\rm Gal}$ vs. $R_{\rm Gal}$), based on the latest {\tt StarHorse} distances \citep{Queiroz2023}. In both panels, the training sample (taken from the APOGEE-{\it Kepler} catalogue of \citealt{Miglio2021}) is overplotted in colours. 
} 
\end{center}
\end{figure}

Using data from the GALAH survey's third data release \citep{Buder2021}, \citet{Hayden2022} have recently demonstrated that it is possible to infer "spectroscopic" (or "chemical") stellar ages for main-sequence turn-off stars with a precision of 1-2 Gyr, using not just one abundance ratio, but a combination of many abundances (not even including carbon and nitrogen), by using supervised machine-learning regression, in particular the popular method extreme gradient-boosted trees ({\tt XGBoost}; \citealt{Chen2016}). A similar exercise using the same technique has been conducted by \citet{He2022}, using red-clump stars observed by the LAMOST survey (which has the advantage of providing also C and N abundances) trained on asteroseismically inferred ages from {\it Kepler}. Their resulting statistical uncertainties, similar to the earlier results for APOGEE \citep{Martig2016, Ness2016}, amount to 31 per cent, which, considering the low resolution of the LAMOST spectra ($R\sim 1\,800$), is remarkable. Similar precisions were obtained earlier by \citet{Mackereth2019} using a Bayesian convolutional neural network and APOGEE DR14 data trained on the APOGEE-{\it Kepler} data \citep{Pinsonneault2018}. Good spectroscopic age estimates have also recently been obtained by \citet{Moya2022} through hierarchical Bayesian modelling for the high-resolution HARPS-GTO sample, by \citet{Bu2020} through Gaussian Process regression on LAMOST data (again using the APOGEE-{\it Kepler} data as a training set), or by \citet{Hasselquist2020} using {\it The Cannon} \citep{Ness2015} for high-luminosity red giants in the Galactic bulge (using the local giant sample of \citealt{Feuillet2018} as a training set).

In this work, we build upon the success of the works of \citet{Hayden2022} and \citet{He2022} and use {\tt XGBoost} to estimate ages for 178\,825 red-giant stars contained in the APOGEE DR17 catalogue \citep{Abdurro'uf2022}. As a training set, we use the carefully produced APOGEE-{\it Kepler} catalogue of \citet{Miglio2021}.

The paper is structured as follows: In Sect. \ref{sec:data} we present the APOGEE data used in this work (including the APOGEE-{\it Kepler} training set). In Sect. \ref{sec:method} we explain the {\tt XGBoost} algorithm. The obtained ages are validated in Sect. \ref{sec:validation}. In Sect. \ref{sec:results}, we show that our ages reproduce known chemical, spatial, and kinematic trends with age, suggesting that our inferred ages are reliable even far from the {\it Kepler} field. Finally, the conclusions are presented in Sect. \ref{sec:conclusions}. Our analysis is reproducible via \url{https://github.com/fjaellet/xgboost_chem_ages}.

\section{Data} \label{sec:data}

The outcome of a machine-learning regressor can only be as reliable as its training set. Therefore, it is often preferable to use a smaller, but statistically significant and highest-quality, dataset for the training phase. 
For red-giant stars, the longest asteroseismic time series have been obtained by the {\it Kepler} satellite \citep{Borucki2010} during its first phase of observations in which it was continuously observing one large sky area in the sky, towards the constellation of Cygnus.

Using APOGEE DR14 \citep{Abolfathi2018} spectroscopic follow-up observations of {\it Kepler} targets \citep{Pinsonneault2018, Yu2018}, \citet{Miglio2021} inferred masses and ages for more than $5\,000$ giants with available Kepler light curves and APOGEE spectra using the Bayesian isochrone-fitting code PARAM \citep{daSilva2006, Rodrigues2017}. This is arguably an asteroseismic+spectroscopic dataset that is of such a high quality that it can be used for training a machine-learning regressor. The median nominal uncertainties of the asteroseismic ages reported in the \citet{Miglio2021} catalogue are 23\% for RGB stars and 10\% for RC stars. For this study we cross-matched the results of \citet{Miglio2021} with the APOGEE DR17 {\tt allStarLite} table \citep{Abdurro'uf2022}, resulting in a preliminary training dataset of $3\,315$ stars. 
We clean the APOGEE-{\it Kepler} sample to contain only stars with APOGEE signal-to-noise ratio ({\rm SNREV}) above 70 and nominal age uncertainties smaller than both 30\%  and 3 Gyr. We also exclude the few apparently young $\alpha$-rich stars (\citealt{Chiappini2015, Martig2015}; here defined as $[\alpha$/Fe] $>0.14 + 0.0001\cdot({\rm age} - 4.1)^5$), since their ages (inferred from single-star evolutionary models) are most probably flawed by binary interactions \citep[e.g.][]{Fuhrmann2017, Jofre2022}.

As we are interested in using the APOGEE abundance measurements as features for the age estimator, we require that the abundance flags\footnote{\url{https://www.sdss4.org/dr17/irspec/abundances/}} for each of the used elements are equal to zero (meaning that their abundance estimates are unproblematic). The chosen chemical features are: $T_{\rm eff}, \log g$, [C/Fe], [CI/Fe], [N/Fe], [O/Fe], [Na/Fe], [Mg/Fe], [Al/Fe], [Si/Fe], [K/Fe], [Ca/Fe], [Ti/Fe], [V/Fe], [Mn/Fe], [Co/Fe], [Ni/Fe], and [Ce/Fe]. The following abundances were discarded: [P/Fe], [S/Fe], [Cr/Fe], [Fe/H], and [Cu/Fe] -- either because there were too few stars with unproblematic ASPCAP flags ([P/Fe] and [Cu/Fe]), because their importance on the final result was negligible ([S/Fe], [Cr/Fe]; see Sect. \ref{sec:shap}), or because we deliberately aim at keeping the age estimate independent from the abundance ratio in question ([Fe/H]). 
This leaves us with $3\,060$ stars covering a metallicity range between $-1$ and $+0.5$. As can be appreciated in Fig. \ref{fig:kiel} (top panel), the training data are also limited to a small range in the spectroscopic Hertzsprung-Russell diagram: $4\,400 \lesssim T_{\rm eff}\ \mathrm{[K]} \lesssim 5\,200$ and $2.2 \lesssim \log g \mathrm{[cgs]} \lesssim 3.4$, compared to the full APOGEE DR17 data. 

When applying our age estimator to the full APOGEE DR17 dataset, we restrict our predictions to the stellar-parameter space covered by the training data (black dots in Fig. \ref{fig:kiel}, top panel), and to [Fe/H] $>-1$. We also require clean abundance flags for all included elements, but allow for lower signal-to-noise ratios ({\tt SNREV} $>50$). The bottom panel of Fig. \ref{fig:kiel} shows the distribution of the training and full sample in Galactocentric cylindrical coordinates, demonstrating the huge gain in area covered with red-giant age estimates -- once our method is applied.

\section{Method} \label{sec:method}

Supervised machine-learning regression models can be trained to fit arbitrarily complex relationships between the introduced data. In our case of tabular data, we want to predict an output column (the so-called "label") based on a set of introduced input parameter columns (so-called "features"). 
The features (in our case stellar parameters and abundances) are the variables of the data that the model uses to train itself to predict the label (stellar age). 
Once the model is trained, it can be used to make predictions (age estimates) based on data for which reliable labels do not yet exist.

\citet{Borisov2021} recently benchmark-tested several regression algorithms for tabular data. They found that the best-performing algorithm, in terms of accuracy and speed, was the well-known tree-based algorithm {\tt XGBoost}. We therefore use this algorithm as the basis for our chemical age estimator in this paper.

\begin{figure}
\begin{center} 
\includegraphics[width=.495\textwidth]{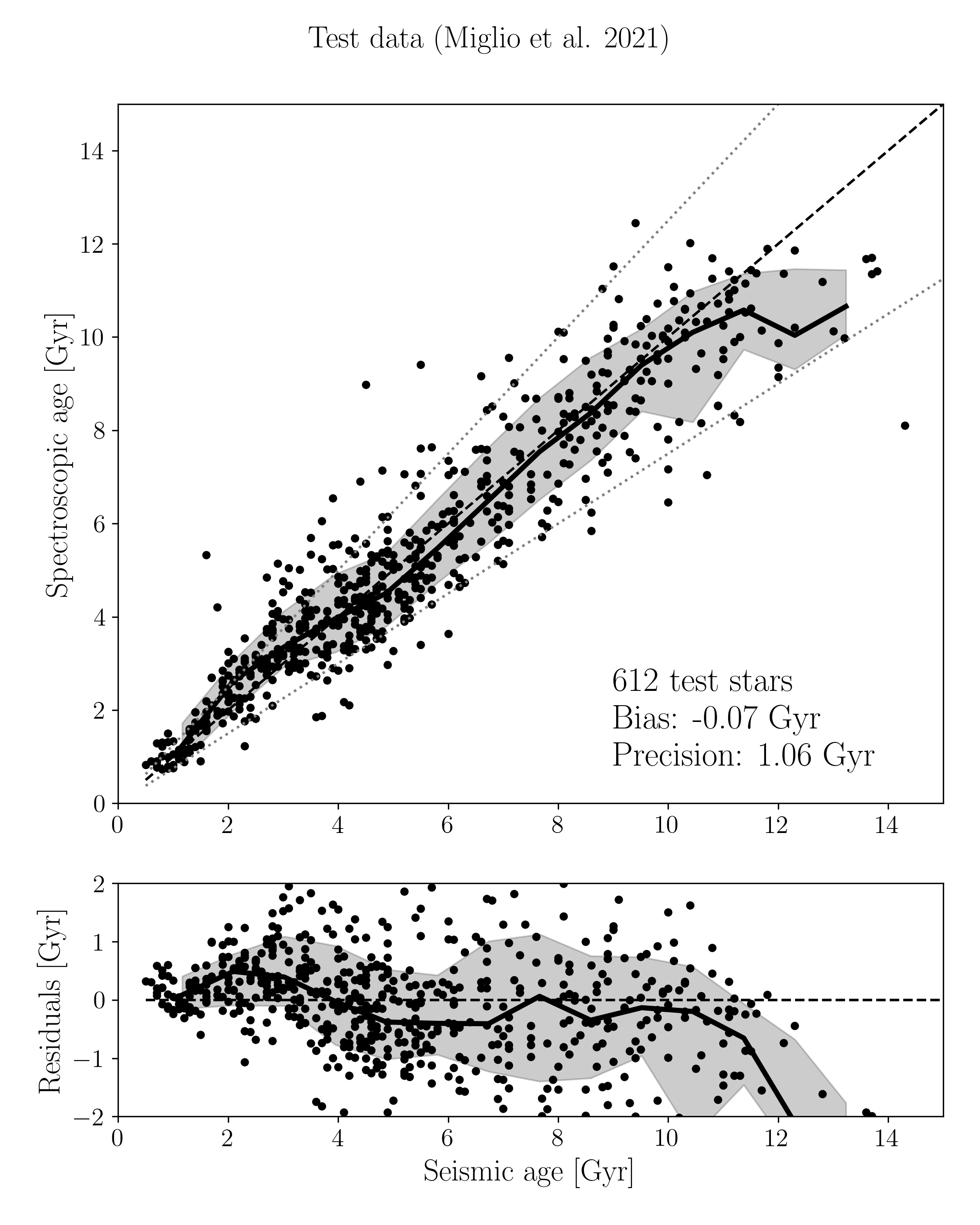}\\
\caption{\label{fig:pred_true} 
Performance of the {\tt XGBoost} chemical age estimator for the unseen test data.  
Top panel: Spectroscopic age vs. asteroseismic age (from \citealt{Miglio2021}), using our default model. The solid black line and the shaded areas delineate the median trend and $1\sigma$ quantiles. 
The dashed line delineates the identity line, while the grey dotted lines correspond to $\pm25\%$ deviation. Bottom panel: Residuals (spectroscopic -- seismic age), showing that our model typically provides reliable ages (with systematic residuals within $\sim0.5$ Gyr for stars with ages $\leq 11$ Gyr. 
} 
\end{center}
\end{figure}

\subsection{XGBoost}

The {\tt XGBoost} algorithm \citep{Chen2016}) is the culmination of previous development in tree-based machine learning. The basis of tree-based algorithms are decision trees, which are graphical representations of possible solutions to a decision based on certain conditions prompted by the values of the input columns (for an introduction in the astronomical context see e.g. \citealt{Ivezic2020}, Chapter 9.7). 

A next level of complexity was achieved by random-forest algorithms \citep{Breiman2001}, which randomly select features (columns) to construct a set of decision trees. On top of random-forest algorithms, one can add so-called gradient-boosting methods, which minimise the errors and increase the performance of the random-forest models. These models use a gradient-descent algorithm to minimise the errors of the sequential models. Finally, in contrast to other gradient-boosted methods, {\tt XGBoost} also performs parallel processing and tree-pruning, handles missing values, and regularises the data to avoid over-fitting.

In consequence, the training time of {\tt XGBoost} is often much smaller compared to other methods with similar predictive power (see \citealt{Borisov2021}). Since the algorithm is very well explained in the original paper \citep{Chen2016} as well as in several online resources, we abstain from a deeper explanation here and only mention that {\tt XGBoost} has a number of hyperparameters that can be tuned to optimise the performance for a given dataset or problem. 

For our use case, we use {\tt scikit-learn}'s \citep{Virtanen2019} 
{\tt GridSearch}\footnote{\url{https://scikit-learn.org/stable/modules/grid_search.html}} to find the optimal hyperparameters for our machine-learning model. The optimal hyperparameters used in our fiducial {\tt XGBoost}\footnote{\url{https://xgboost.readthedocs.io/en/stable/}} regressor are: {\tt learning\_rate}$=0.005$, {\tt max\_depth}$=7$, {\tt min\_child\_weight}$=10$, {\tt n\_estimators}$=1500$, {\tt subsample}$=0.6$. The performance of the default model is shown in Fig. \ref{fig:pred_true} and discussed in detail in Sect. \ref{sec:uncerts}. In addition, we also provide alternative estimates with individual uncertainties, by using quantile regression (available in the latest version of {\tt XGBoost}) to include the prediction of $1\sigma$ confidence intervals in the inference (see Appendix \ref{sec:datamodel}). In this paper we mainly use the results of our fiducial {\tt XGBoost} run (column {\tt spec\_age\_xgb} in Table \ref{datamodel}).

\subsection{SHAP values}\label{sec:shap}

The concept of SHAP (SHapley Additive exPlanations) values provides an elegant way to understand the output of a machine learning model \citep{Lundberg2017}. In the case of {\tt XGBoost}, they can be used to understand how each feature has an impact on the predictions of the model. The sum of SHAP values is equal to the difference between the expected output and the baseline output. In other words, a positive value means that the feature increases the output’s model, while a negative value decreases it. For our case, the SHAP values have been used not only to recognize the features that have the greatest impact on the model, but also to discard those that do not contribute to the prediction (the APOGEE DR17 elemental abundances discarded in Sect. \ref{sec:data}). 

\begin{figure}
\begin{center} 
\includegraphics[width=.495\textwidth]{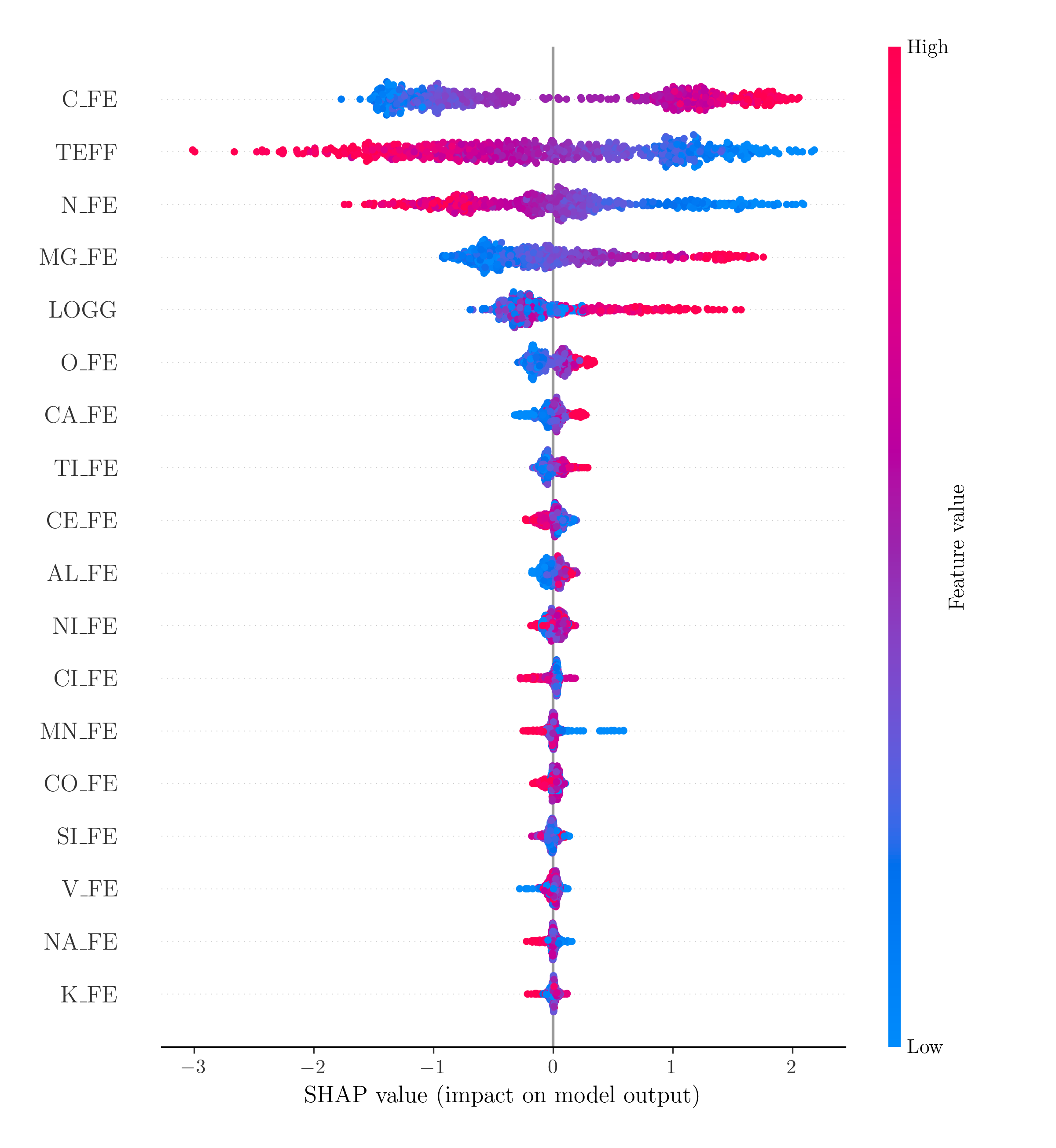}\\
\caption{\label{fig:shap} 
SHAP "bee-swarm" plot\protect\footnotemark. The rows correspond to the input features, ordered by importance. In each row, one dot corresponds to a star in the test dataset, coloured by its (normalised) feature value. The position of each dot indicates how much and in which way each feature contributes to its output label (age). 
}
\end{center}
\end{figure}

Figure \ref{fig:shap} demonstrates that the most influential features in the age prediction are the effective temperature, $T_{\rm eff}$, the chemical abundances [C/Fe], [Mg/Fe], [N/Fe], and the surface gravity, $\log g$. In addition, Fig. \ref{fig:shap} shows the exact effect each of the features has on the estimated age:
for example, for the case of effective temperature (redder dots in the first line), we observe that an increase in its value implies a negative SHAP value, i.e. a decrease in the inferred age. 
The usefulness of carbon and nitrogen abundances for the prediction of red-giant ages in the absence of asteroseismology is discussed in detail by \citet{Martig2016}. These authors performed a polynomial feature regression on the APOGEE DR12 labels \{$T_{\rm eff}, \log g$, [M/H], [C/M], [N/M], [C+N/M]\} to the APOKASC-1 ages \citep{Pinsonneault2014} and obtained reasonably precise age estimates for 52\,000 APOGEE stars ($\sim 40\%$). Similar precision was achieved by \citet{Ness2016}, using the same training data as \citet{Martig2016}, but a different technique (The Cannon, a data-driven regression technique directly trained on the spectra; \citealt{Ness2015}). 

\footnotetext{\url{https://shap.readthedocs.io/en/latest/example_notebooks/api_examples/plots/beeswarm.html}}
The fact that [C/Fe], $T_{\rm eff}$, [N/Fe], [Mg/Fe], and $\log g$ are the most important features for age determination is very much in line with the findings of \citet{Bu2020} who, using a variety of machine-learning regression techniques, found that these features contained most information for predicting ages of LAMOST stars (in the absence of reliable mass estimates). Also the work of \citet{Ciuca2022} used these five features (+ [Fe/H]) to determine ages for APOGEE stars. We also see from Fig. \ref{fig:shap} that most other individual abundance ratios (first [O/Fe] and other $\alpha$ elements, then [Ce/Fe], and then iron-peak elements) have a smaller impact on the estimated age. Although all abundance measurements are used by {\tt XGBoost} to some extent to estimate an even more precise age, the precision improvement of the full model using all abundances over a model that only uses the feature space \{[C/Fe], $T_{\rm eff}$, [N/Fe], [Mg/Fe], $\log g$\} is only $\sim 10\%$. This suggests that while each element is unique, the key information is stored in only a few vectors \citep[in agreement with previous works, e.g.][]{PriceJones2018, Weinberg2019, Ratcliffe2020}. Interestingly, Ce is not among the most influential features despite being an s-process element for which the correlation with age is strong \citep[e.g.][]{Maiorca2011, Spina2018, Magrini2018, DelgadoMena2019, Casamiquela2021, Sales-Silva2022, Casali2023}. This might be due to the still sizeable uncertainties of this element in the APOGEE DR17 catalogue (see also \citealt{Hayes2022}).

\subsection{Biases and uncertainties}\label{sec:uncerts}

The performance of our default model (trained on 80\% of the \citealt{Miglio2021} data) is shown for the test dataset (the 20\% of the data not used in the training phase) in Fig. \ref{fig:pred_true}. We observe a very clear linear trend with little wiggles ($\lesssim0.5$ Gyr for almost the full age range), and a determination coefficient of $R^2 = 0.91$. Our spectroscopic ages tend to slightly overestimate (by $<0.5$ Gyr) ages for stars around 2 Gyr, while for stars between 5 and 7 Gyr our age scale tends to underestimate the seismic age by similar amounts. Beyond seismic ages of $\sim 11$ Gyr, our method provides more seriously underestimated ages compared to the seismic scale. 

Considering that the age uncertainties associated with the training data are of order 10\% for RC and 25\% for first-ascent red-giant branch (RGB) stars\footnote{We note that these dispersions do not reflect the systematic uncertainties which are considerably higher for red-clump stars (e.g. \citealt{Casagrande2016, Anders2017}) due to the poorly constrained amount of mass loss after the first dredge-up phase.}, the amount of scatter around the one-to-one relation in Fig. \ref{fig:pred_true} is indeed surprisingly small (almost exactly the same values), indicating that the regressor is well trained and that most of the variance stems from the training set itself. In addition, and contrary to the training set, {\tt XGBoost} is not extrapolating the age scale beyond the age of the Universe (13.7 Gyr), which again is surprising. 

\begin{figure}
\begin{center} 
\includegraphics[width=.495\textwidth]{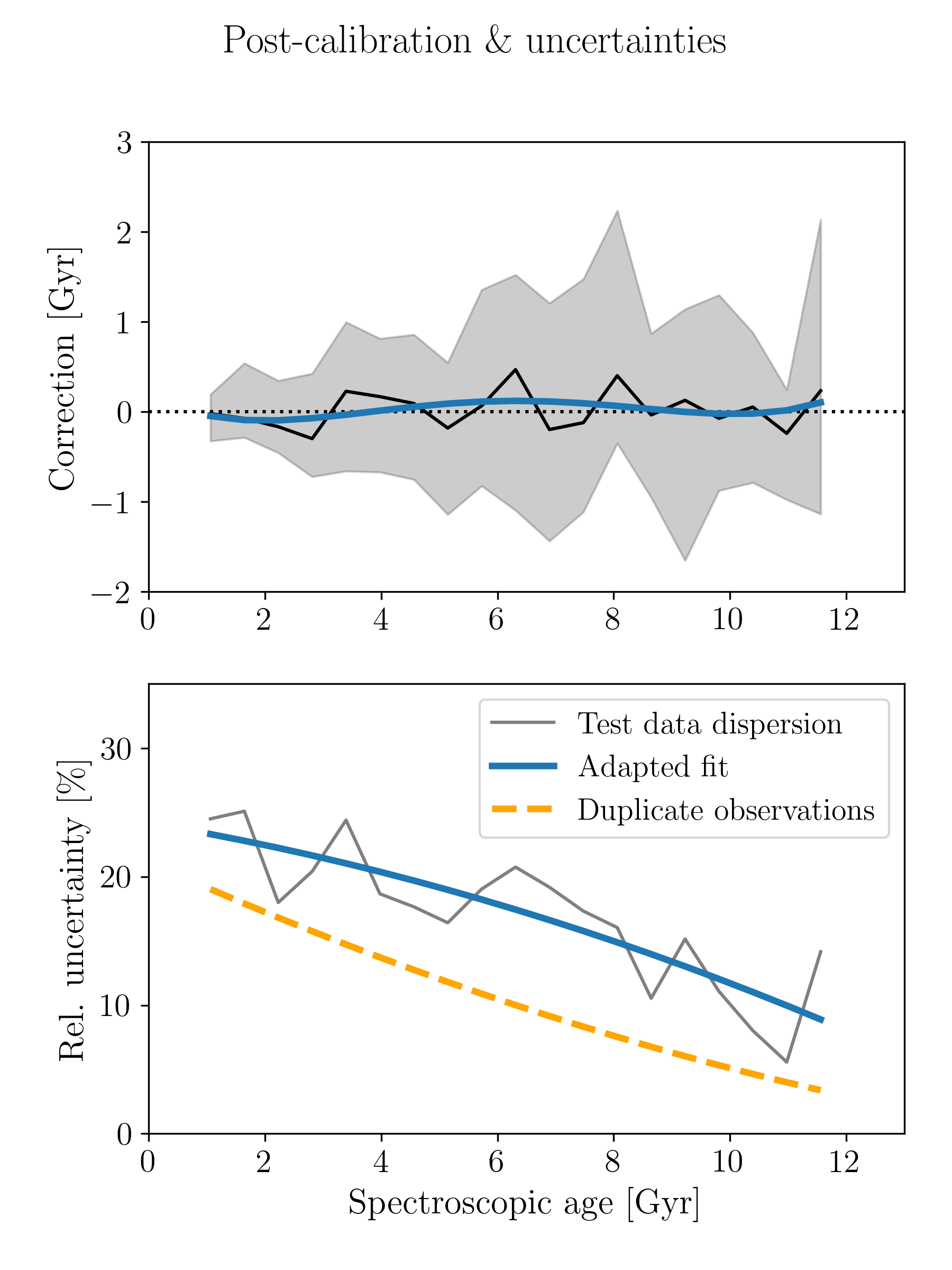}\\
\caption{\label{fig:uncerts} 
Possible a-posteriori calibration of the age estimates (top panel) and estimation of the internal statistical uncertainties (bottom panel) using the same data as in Fig. \ref{fig:pred_true}. The top panel shows the median age bias (test seismic age $-$ spectroscopic age) as a function of the spectroscopic age. The bottom panel shows the relative age error (width of the grey distribution in the top panel normalised by the spectroscopic age). As a sanity check, the orange dotted line in the bottom panel shows the median relative age dispersion among duplicate APOGEE observations. In both panels, we also show the simple polynomial fits used to characterise the bias and $1\sigma$ age uncertainty in our catalogue.
}
\end{center}
\end{figure}

We also tested training separate {\tt XGBoost} regressors for the RC and RGB stars, respectively. However, a global increase in performance could not be appreciated, most probably because the default algorithm also learns the separation between RC and RGB stars from the data (in particular, $T_{\rm eff}$, $\log g$, [C/Fe], and [N/Fe]). Our default {\tt XGBoost} model is therefore capable of estimating meaningful ages for both RC and RGB stars. We appreciate that the performance for the RC stars is significantly better (lower dispersion around the identity line), but the model still works well on average for RGB stars. The statistical age uncertainty of our method is lower than 25\% and only weakly dependent on the estimated stellar age (see Fig. \ref{fig:uncerts}, bottom panel).

\section{Validation}\label{sec:validation}

Age estimates for field stars are, as explained in the Introduction, highly dependent on the accuracy of stellar evolutionary models, even in the case of the highest quality data. On top of the systematic age biases that are unavoidable and very hard to quantify especially in the case of red-giant stars (related to e.g. assumptions about mass loss, mixing, rotation, etc.; \citealt{Noels2015}), there are also sizeable statistical uncertainties. Nevertheless, comparisons to other model-dependent age estimates are of prime importance to estimate the amount of systematic and statistical uncertainties.
In this section, we infer internal uncertainties of our method and compare the resulting ages to asteroseismic ages obtained from observing campaigns other than {\it Kepler}, open cluster ages, and APOGEE DR17 age estimates from other machine-learning methods. We provide additional verification of our ages against isochrone fitting and an empirical [C/N] calibration in Appendix \ref{sec:appendix}.

\begin{figure}
\begin{center} 
\includegraphics[width=.495\textwidth]{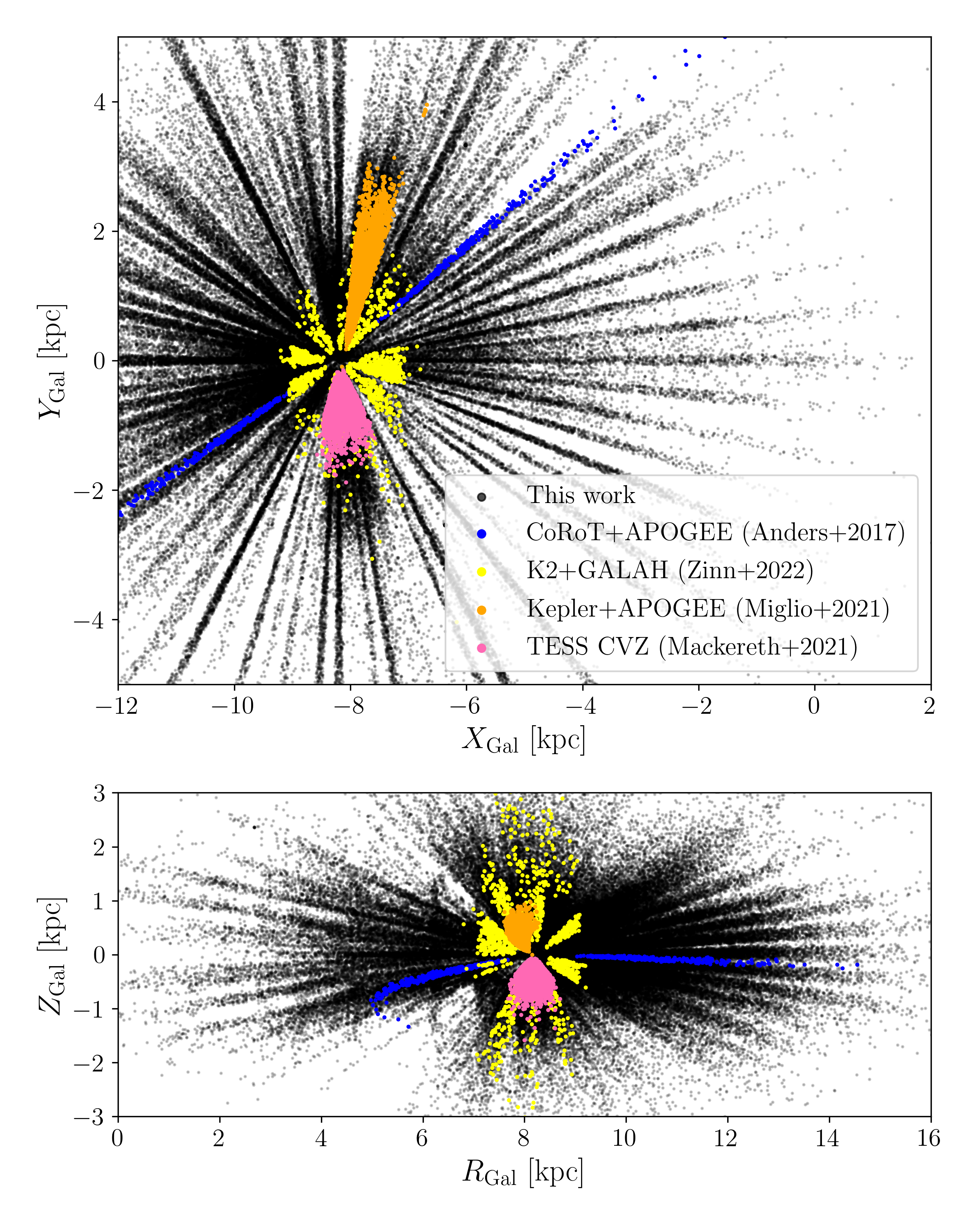}\\
\caption{\label{fig:xyrz} 
Galactic distribution of asteroseismic+spectroscopic red-giant samples, superimposed on our sample of APOGEE DR17 red-giants with spectroscopic ages determined in this work. 
}
\end{center}
\end{figure}

\subsection{Duplicate APOGEE observations}

More than 11\,000 stars in our selected region of the {\it Kiel} diagram (Fig. \ref{fig:kiel}) are contained more than once in APOGEE DR17 catalogue (for example, stars that have been observed in different fields or even by both the SDSS Telescope and the du Pont Telescope). These duplicate observations can serve to estimate realistic internal uncertainties for stellar parameters and elemental abundances (e.g. \citealt{Jonsson2020}, Tables 10 \& 11). In our case we can use these multiple observations also to assess the uncertainties of our age estimates: the scatter in the stellar parameters and abundances naturally leads to a scatter in the inferred ages. The results of this experiment are shown in the bottom panel of Fig. \ref{fig:uncerts} (orange dashed curve). Reassuringly, the dispersion among multiple observations (with often significantly different signal-to-noise ratios and, therefore, elemental-abundance precisions) is always lower than the uncertainty estimated from the test dataset.

\begin{figure*}
\begin{center} 
\includegraphics[width=.49\textwidth]{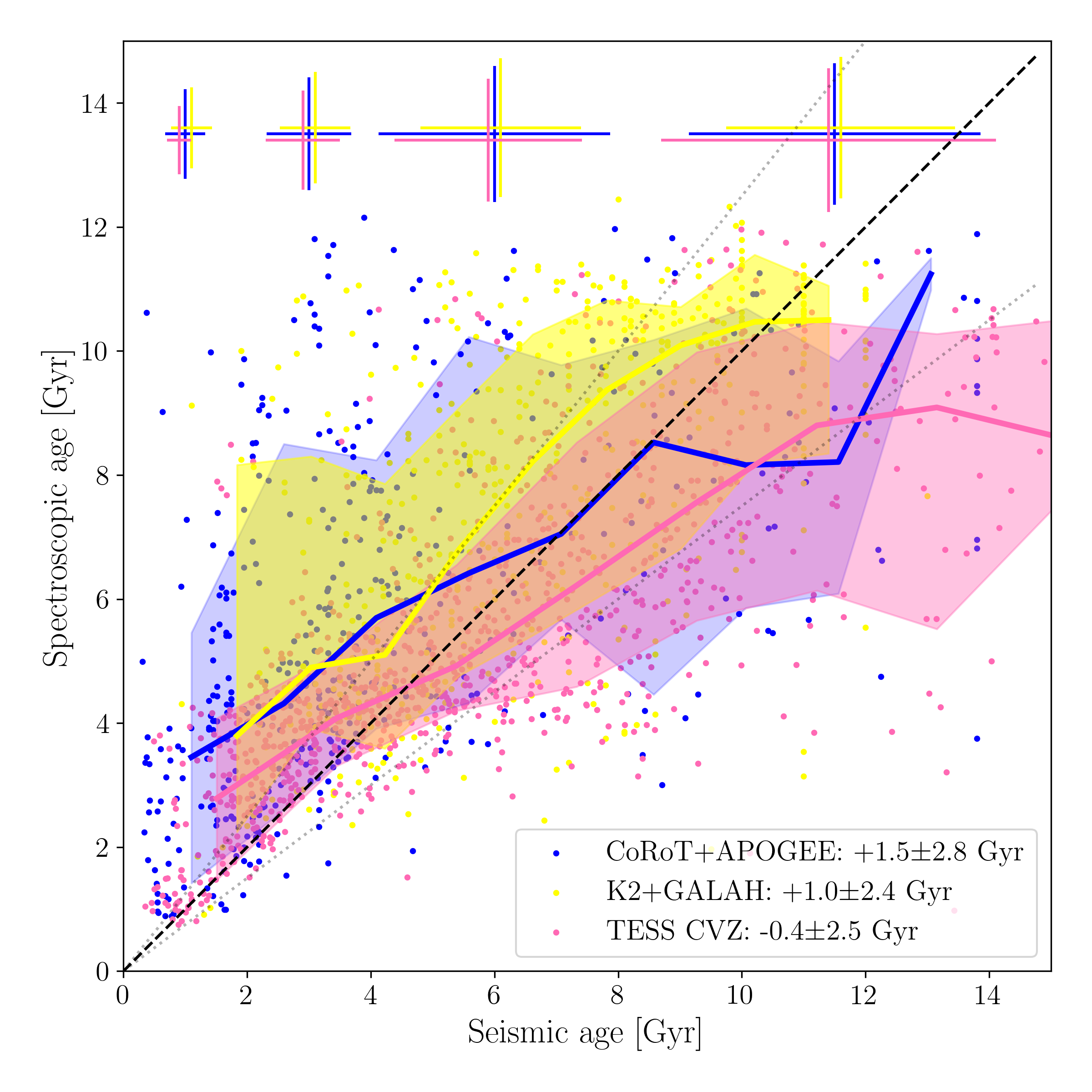}
\includegraphics[width=.49\textwidth]{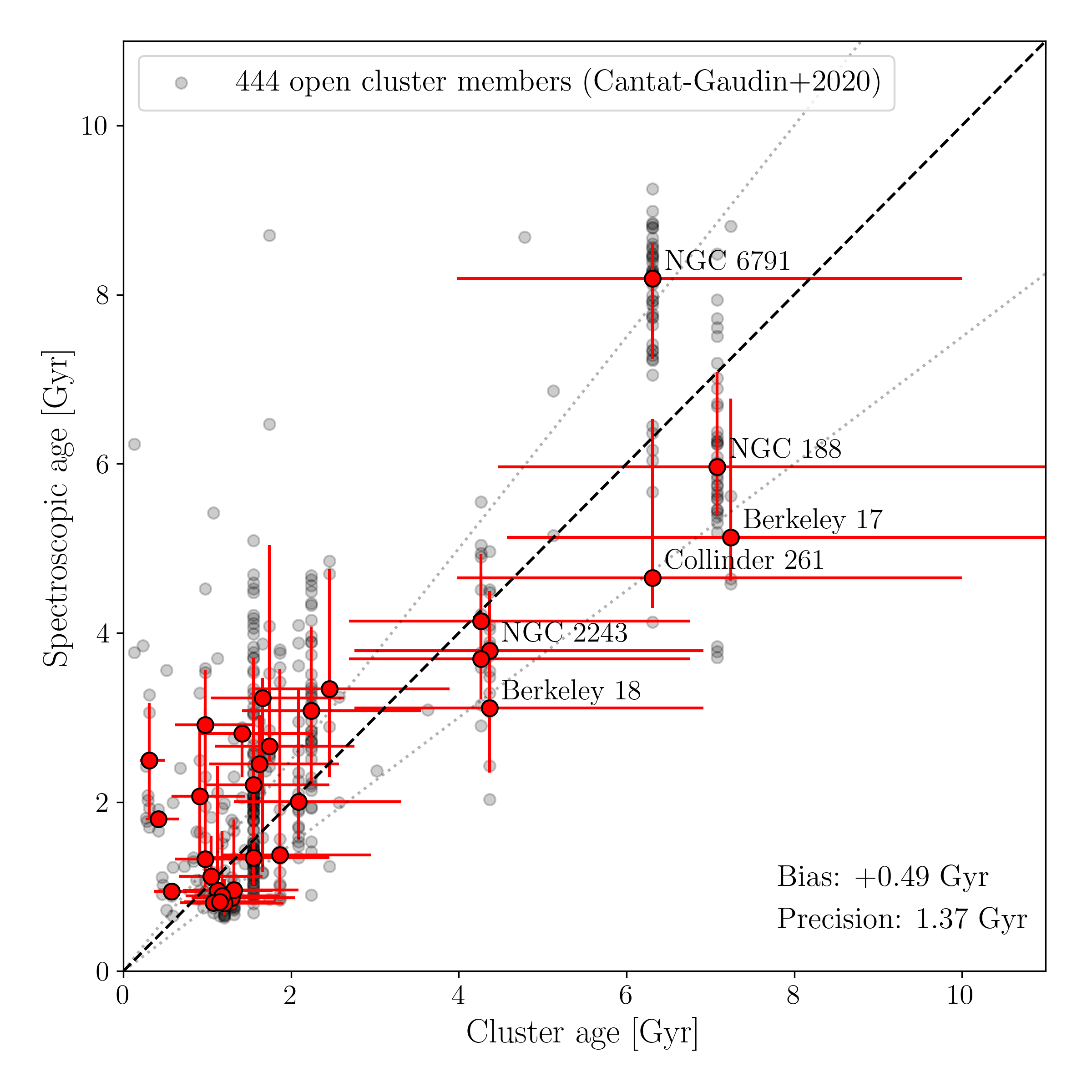}\\
\includegraphics[width=.33\textwidth]{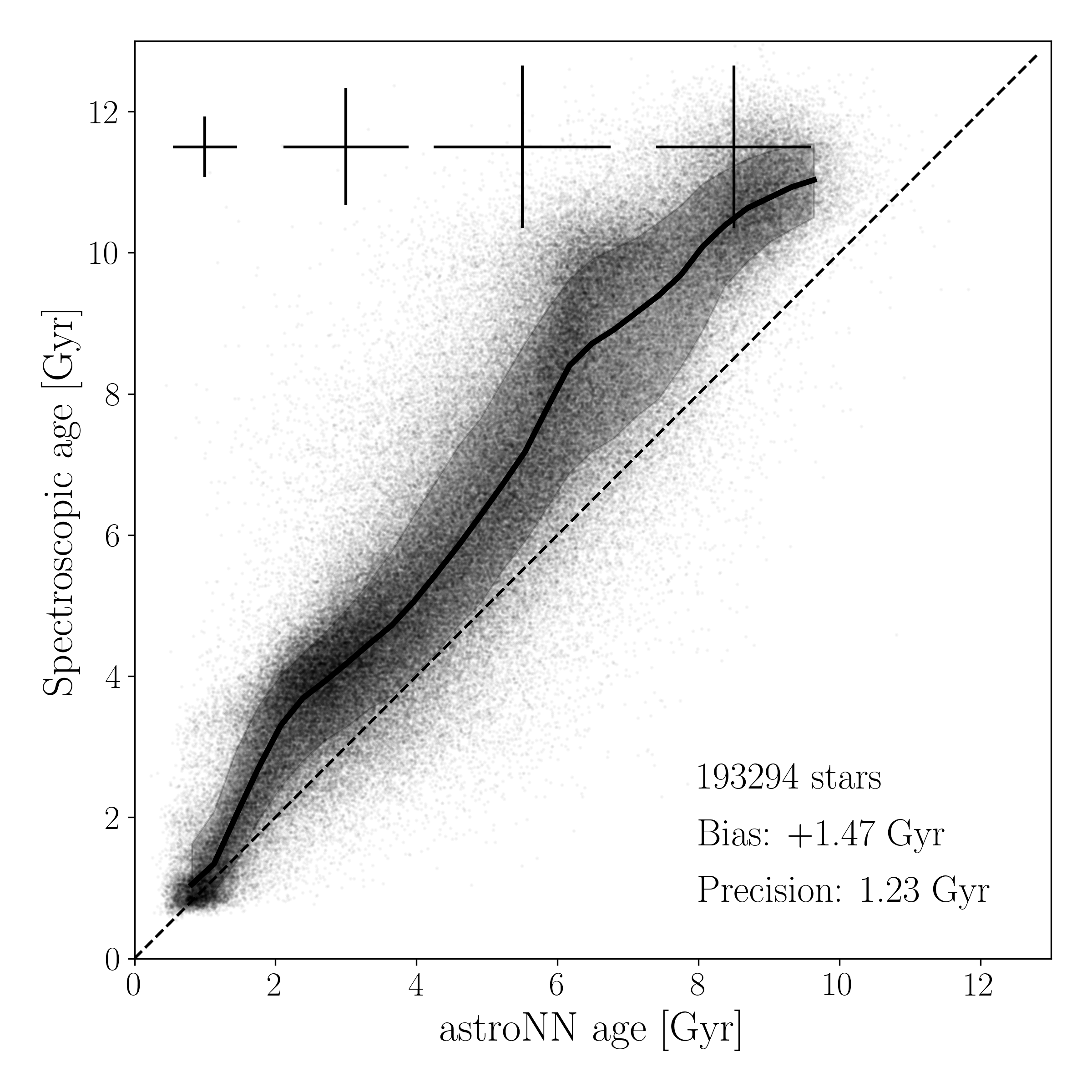}
\includegraphics[width=.33\textwidth]{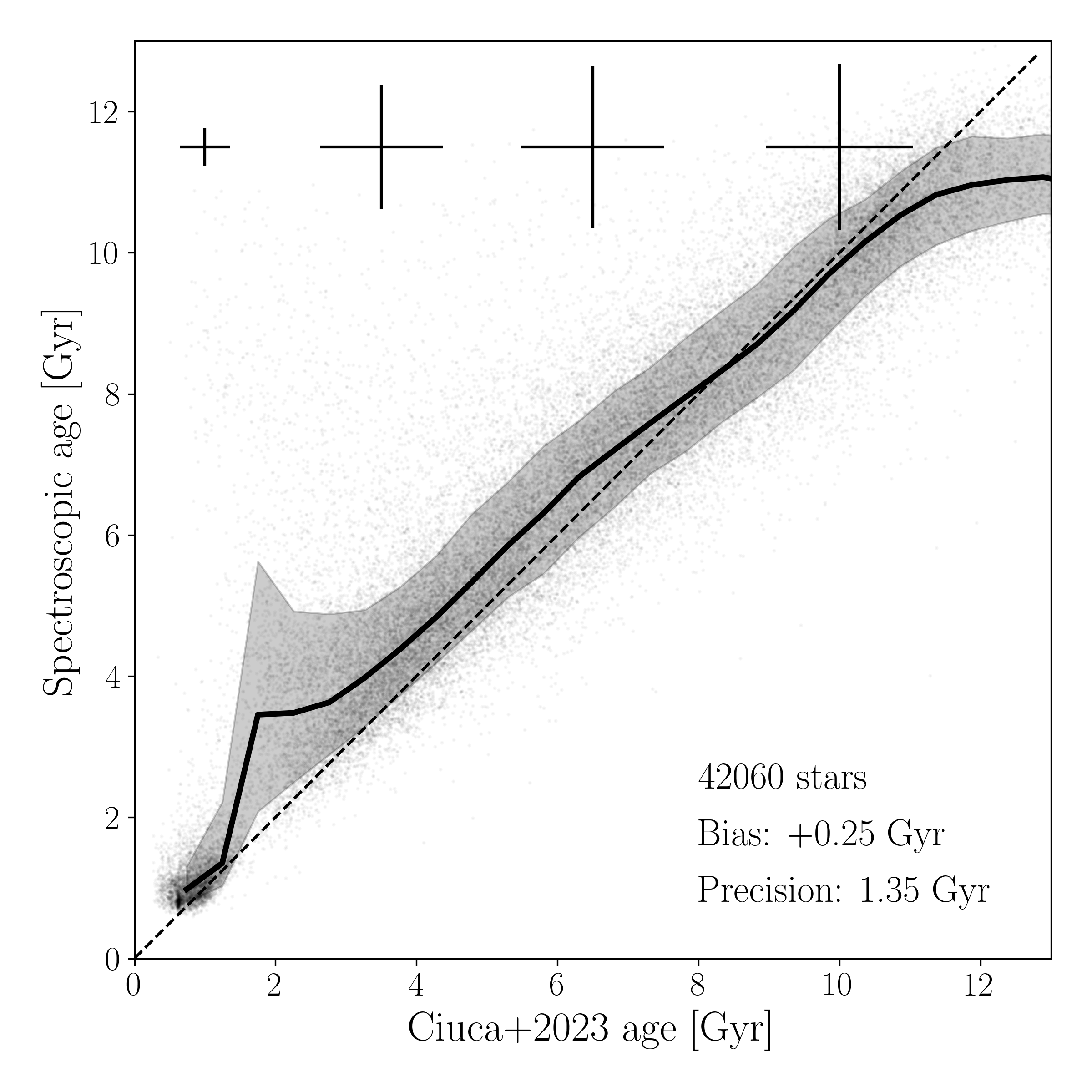}
\includegraphics[width=.33\textwidth]{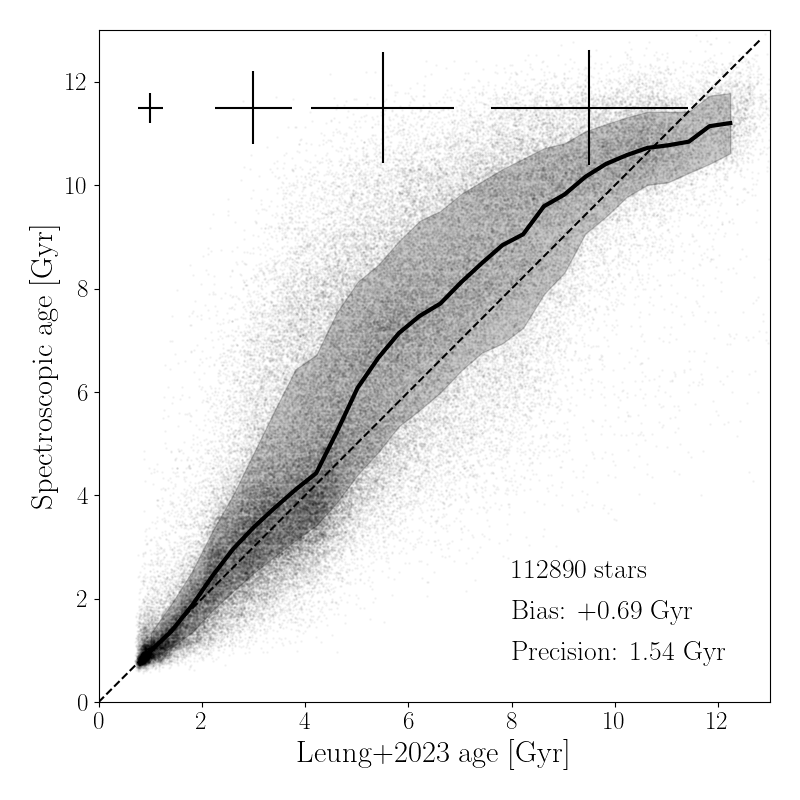}
\caption{\label{fig:age_comp_5} 
Comparisons of our uncalibrated age estimates with ages from the literature. The lines and shaded bands in each panel delineate the running median and $1\sigma$ quantiles, respectively. 
Top left panel: age estimates from recent asteroseismic catalogues (obtained from shorter time series than the {\it Kepler} data), as indicated in the legend. Top right: open cluster ages as listed in the catalogue of \citet{Cantat-Gaudin2020}. Grey dots correspond to individual cluster members (with membership probabilities $>0.95$), while the red errorbars show the median statistics in each cluster with more than two members (the ones older than 4 Gyr are annotated). Bottom row: Machine-learning field star estimates for APOGEE DR17 using the same training dataset as used in this paper \citep{Miglio2021}. Bottom left: DR17 {\tt astroNN} ages \citep{Leung2019}. Bottom middle: Age estimates from \citet{Ciuca2022}. Bottom right: Age estimates obtained by \citet{Leung2023} using a random-forest regressor trained on the APOGEE DR17 spectral latent space.  
}
\end{center}
\end{figure*}

\subsection{Asteroseismic ages in other fields (CoRoT, K2, TESS)}

Asteroseismology coupled with spectroscopy can significantly reduce the uncertainties of red-giant ages \citep[e.g.][]{Miglio2013, Pinsonneault2014}. This is exactly the reason why we have chosen the APOGEE-{\it Kepler} dataset of \citet{Miglio2021} as a training set for our method. The concordance with this age scale is inherent to the machine-learning method used in this paper, and was demonstrated in Fig. \ref{fig:pred_true} for the unseen test data.

Here we compare our age estimates for the full APOGEE DR17 data with other age estimates that used similar techniques as \citet{Miglio2021}, but on other asteroseismic and spectroscopic data. The main reason is that we need to assess whether the stellar populations contained in the training set (i.e. the {\it Kepler} field) are representative enough - so that our method can also successfully be used to derive ages for stars in other parts of the Galaxy. 
To verify this, we study the CoRoT-APOGEE data presented in \citet{Anders2017}, the K2-GALAH data \citep{Zinn2022}, and the first Galactic Archaeology data from the Southern continuous viewing zone of the TESS satellite (TESS CVZ; \citealt{Mackereth2021}). The Galactic distribution of all these targets, along with our APOGEE DR17 stars with spectroscopic age estimates, is shown in Fig. \ref{fig:xyrz}. The distances for the \citealt{Zinn2022} and for the \citet{Mackereth2021} stars in this plot were taken from \citet{Queiroz2023} and \citet{Anders2022}, respectively. 
 
Figure \ref{fig:age_comp_5} (top left panel) presents the direct comparison with the asteroseismic ages obtained by \citet{Anders2017, Zinn2022}, and \citet{Mackereth2021}. The plot shares some similarities with the comparison plot for test sample shown in Fig. \ref{fig:pred_true}: in the youngish regime, the {\tt XGBoost} ages appear overestimated with respect to the seismic scale for all three external datasets, albeit much more pronounced than in the {\it Kepler} sample. In the intermediate regime (between 4 and 10 Gyr), the concordance is better on average, and there is no consistent trend among the three external seismic samples (in this regime, the age scale of \citealt{Zinn2022} is lower, while the one of \citealt{Mackereth2021} is higher than ours). For the oldest ages ($>10$ Gyr), the {\tt XGBoost} ages appear underestimated with respect to the seismic scale, at least for the case of the TESS-CVZ sample.

Finally, it is important to recall that also the asteroseismic+spectroscopic catalogues used for comparison here are prone to sizeable statistical and systematic uncertainties. Due to the shorter time series of especially CoRoT and K2, these uncertainties are typically much larger than the ones associated to the {\it Kepler} training data. To illustrate this, the typical statistical age uncertainties for each of the considered catalogues are shown as error bars in Fig. \ref{fig:age_comp_5}. 

\subsection{Open clusters \citep{Cantat-Gaudin2020}}

Figure \ref{fig:age_comp_5} (top right panel) shows the comparison of our spectroscopic age estimates with open-cluster (OC) ages obtained by \citet{Cantat-Gaudin2020} by {\it Gaia} DR2 colour-magnitude diagram fitting with an artificial neural network trained on a mixture of simulations and real data (mostly stemming from \citealt{Bossini2019}). As indicated by the error bars in the figure, the OC age estimates also come with sizeable uncertainties. Keeping this in mind, the concordance between our age estimates and the OC ages is not too bad: on average, our ages are greater than the OC ages by 0.48 Gyr, and the dispersion is 1.37 Gyr ($\sim 40\%$ larger than the uncertainties estimated from the test dataset in Fig. \ref{fig:pred_true}). 

In spite of the statistical concordance, however, we observe a tendency to overestimate age for a group of clusters younger than 2.5 Gyr. \citet{Queiroz2023} observed a similar behaviour in their comparison of sub-giant star age estimates with the OC age scale of \citet{Cantat-Gaudin2020}. This suggests that either the OC age scale of \citet{Cantat-Gaudin2020} has to be revised (perhaps by enlarging the OC training set with reliable ages, or by better taking into account metallicity effects) or that both the isochrone ages of \citet{Queiroz2023} and our {\tt XGBoost} ages suffer from similar (but independent) systematics. Part of these systematics are caused by the coldest members of young clusters (i.e. the largest giants), which is in part expected, since the spectroscopic abundances of young stars ($<500$ Myr) can be significantly biased (see e.g. \citealt{Spina2022}, Sect. 5.6).

\subsection{Machine-learning age estimates for APOGEE DR17} 

The lower panels of Fig. \ref{fig:age_comp_5} show comparisons of our age estimates with three independent attempts at producing field-star age estimates for APOGEE DR17 red-giant stars, also using machine-learning techniques and APOGEE-{\it Kepler} data as their training set. In all panels of Fig. \ref{fig:age_comp_5} we show comparisons to our uncalibrated age scale (the differences between calibrated and uncalibrated ages are marginal).

\citet{Leung2019} presented a neural-network algorithm, {\tt astroNN}, that is capable of determining stellar atmospheric parameters, elemental abundances, and additional desirables such as distance and age from APOGEE spectra. The DR17 {\tt astroNN} catalogue is available at the SDSS DR17 website\footnote{\url{https://www.sdss4.org/dr17/data_access/value-added-catalogs}}.

The age estimation of {\tt astroNN} follows the procedure of \citet{Mackereth2019} and uses a similar training set as ours. Instead of the \citet{Miglio2021} ages, however, the {\tt astroNN} ages are based on the APOKASC-2 catalogue \citep{Pinsonneault2018}, complemented with the 96 low-metallicity stars of \citet{Montalban2021}. 
In view of the similarity of the training sets (and the overall philosophy of obtaining age estimates from spectroscopy alone), it is not surprising that the comparison shown in Fig. \ref{fig:age_comp_5} (lower left panel) shows much less scatter, albeit with a significant systematic trend (which is most likely inherited from the different age scale in their training set): except for the youngest stars, our ages are typically larger than the {\tt astroNN} estimates by a factor of $\approx1.2$. This, however, is reassuring, since the age plateau present in the {\tt astroNN} catalogue at $\lesssim 9$ Gyr is inconsistent with the age of the Galactic disc (which is $>12$ Gyr; e.g. \citealt{Fuhrmann2017a, Rendle2019}).

In a different attempt at determining age estimates for APOGEE DR17 red giants (but also using the data of \citealt{Miglio2021} as a training set), \citet{Ciuca2022} used a Bayesian neural-network architecture, BINGO \citep{Ciuca2021}, to infer precise stellar age estimates for 68\,360 stars with exquisite APOGEE signal-to-noise ratios ({\tt SNREV} $>100$). As input, \citet{Ciuca2022} use the stellar parameters $T_{\rm eff}$ and $\log g$ as well as the abundances [Fe/H], [Mg/Fe], [C/Fe] and [N/Fe], each with their associated uncertainties. Following the method of \citet{Das2019}, the hidden parameters (a.k.a. weights and biases) of a neural network with two fully-connected layers were optimised during the training phase. Then, by marginalising over the posterior distribution of the neural network parameters, a posterior age distribution for each individual star was obtained.

The lower middle panel of Fig. \ref{fig:age_comp_5} shows the comparison of our spectroscopic age estimates with the ages obtained by \citet{Ciuca2022}. Except for the small jump at around $\approx1.8$ Gyr (which is partly due to low statistics in that age regime), we find a remarkable concordance between the two age scales, with little systematic differences (+0.24 Gyr) and scatter (1.34 Gyr). The formal uncertainties of the BINGO ages are also similar to ours.

Very recently, \citet{Leung2023} presented a new catalogue of APOGEE DR17 age estimates that addresses some of the issues present in the {\tt astroNN} catalogue. In particular, they present the implementation of a novel technique that they dub "variational encoder-decoder", which reduces the dimensionality of the APOGEE spectra to a four-dimensional latent space that does not contain strong elemental-abundance information. In a second step, \citet{Leung2023} trained a random-forest regressor to learn the relationship between the latent-space parameters and stellar ages, using the APOGEE-{\it Kepler} data of \citet{Miglio2021}. They found that this method delivers more precise and accurate results than their earlier {\tt astroNN} attempt, delivering age uncertainties of $\sim 22\%$. Perhaps the main advantage of their method over other attempts (including ours) is that their ages do not depend on [Mg/Fe] (see Sect. \ref{sec:caveats}).

The lower right panel of Fig. \ref{fig:age_comp_5} compares our spectroscopic age estimates with the estimates of \citet{Leung2023}. In the young regime ($<4$ Gyr), we find very good agreement and little dispersion between the two age scales. For the intermediate and old regime, both systematic differences and dispersion increase, so that the overall concordance between the two methods is worse than in the case of \citet{Ciuca2022}, and we also see more dispersion than in the comparison to the {\tt astroNN} ages. This is to some degree expected, because the method of \citet{Leung2023} explicitly excludes the use of part of the information contained in the stellar spectrum (also indicated by the larger uncertainties; see error bars in Fig. \ref{fig:age_comp_5}). Nevertheless, the comparison is reassuring, in particular with regards to the younger stars where we see very little systematic trends.

\subsection{Caveats}\label{sec:caveats}

Some caveats have to be taken into account when using our age estimates. Most importantly, the age estimates critically depend on the quality of the APOGEE DR17 atmospheric parameters and abundances. Since our fiducial method does not provide individual uncertainty estimates, the user may decide whether to apply further quality cuts to the ones already imposed.

We are extrapolating the relationships found in the {\it Kepler} data to the whole Galactic volume covered by the APOGEE DR17 red-giant sample. This could potentially be dangerous as well, and it is to some degree surprising that it works so well. The reason for that is most certainly stellar migration: the stellar population of the {\it Kepler} field has such a great variety in ages and abundances because its stars came from vastly different birth positions in the Galactic disc (see e.g. \citealt{Minchev2013, Lagarde2021, Miglio2021, Ratcliffe2023}), so that this caveat is milder than it could have been.

By construction, our age estimates are not independent of chemical abundances (most importantly, [C/Fe], [N/Fe], and [Mg/Fe]). This tendentially makes inferences of the age-abundance relations slightly circular, at least for the elemental abundance ratios that are most important for the {\tt XGBoost} model. 

Despite the large sample size and coverage of the Galactic disc, our sample is affected by some important selection effects. Most importantly, it is restricted to stars with [Fe/H] $>-1$, and therefore to disc stars. A detailed comparison of our sample with Milky Way models will require careful forward modelling to take into account these selection effects.
    
While traditional chemical clocks such as [$\alpha$/Fe] or [Y/Mg] should be quite robust also for close binaries, post-common-envelope phase binaries, and binary merger products, our age estimates also rely on stellar parameters and chemical abundances that are heavily affected by non-binary evolution. This is the reason why, despite the fact that we eliminated the conspicuous "young $\alpha$-rich" stars (a.k.a [$\alpha$/Fe]-enhanced over-massive stars) from the training set, we still find some of them in the full APOGEE sample. In particular, we find 706 stars (0.4\% of the total sample) with [Mg/Fe]$>0.2$ and spectroscopic ages $\lesssim6$ Gyr. These stars are very likely products of binary evolution; their ages should not be used. Another small part of our dataset (561 stars) that is most likely affected by effects of binary evolution are rapidly rotating ($v \sin i > 10$ km/s) red giants - they were recently analysed using APOGEE DR16 by \citet{Patton2023}. We include flags for both these categories in our results, along with flags for stars that are slightly bluer (4\,137 stars) or redder (280 stars) than the bulk of the training set (and that thus may have less reliable ages). There are 5\,383 flagged stars in total.

\section{Results}\label{sec:results}

We applied our {\tt XGBoost} model to $193\,478$ APOGEE DR17 stars with {\tt SNREV}$>50$, [Fe/H]$>-1$, and clean abundance flags (see Sect. \ref{sec:data}) located in the small $T_{\rm eff} - \log g$ box highlighted in Fig. \ref{fig:kiel}, thus avoiding extrapolation of our results into a regime of stellar parameters not covered by the training set. After cleaning for duplicate {\tt APOGEE\_ID}s (keeping the results with the highest {\tt SNREV}), we provide spectroscopic age estimates for 178\,825 unique APOGEE DR17 stars. To further ensure that these results are meaningful, we show a number of validation plots that reproduce well the expected chemical, positional, and kinematic trends with age. When interpreting these plots, we recall that our absolute age scale is likely subject to some systematics, as illustrated in Sect. \ref{sec:validation}. For a detailed analysis of the chemical evolution of the Galactic disc with this sample, including a proper treatment of stellar radial migration, we refer the reader to \citet{Ratcliffe2023}.

\subsection{Age-metallicity relation}\label{sec:amr}

\begin{figure}
\begin{center} 
\includegraphics[width=.495\textwidth]{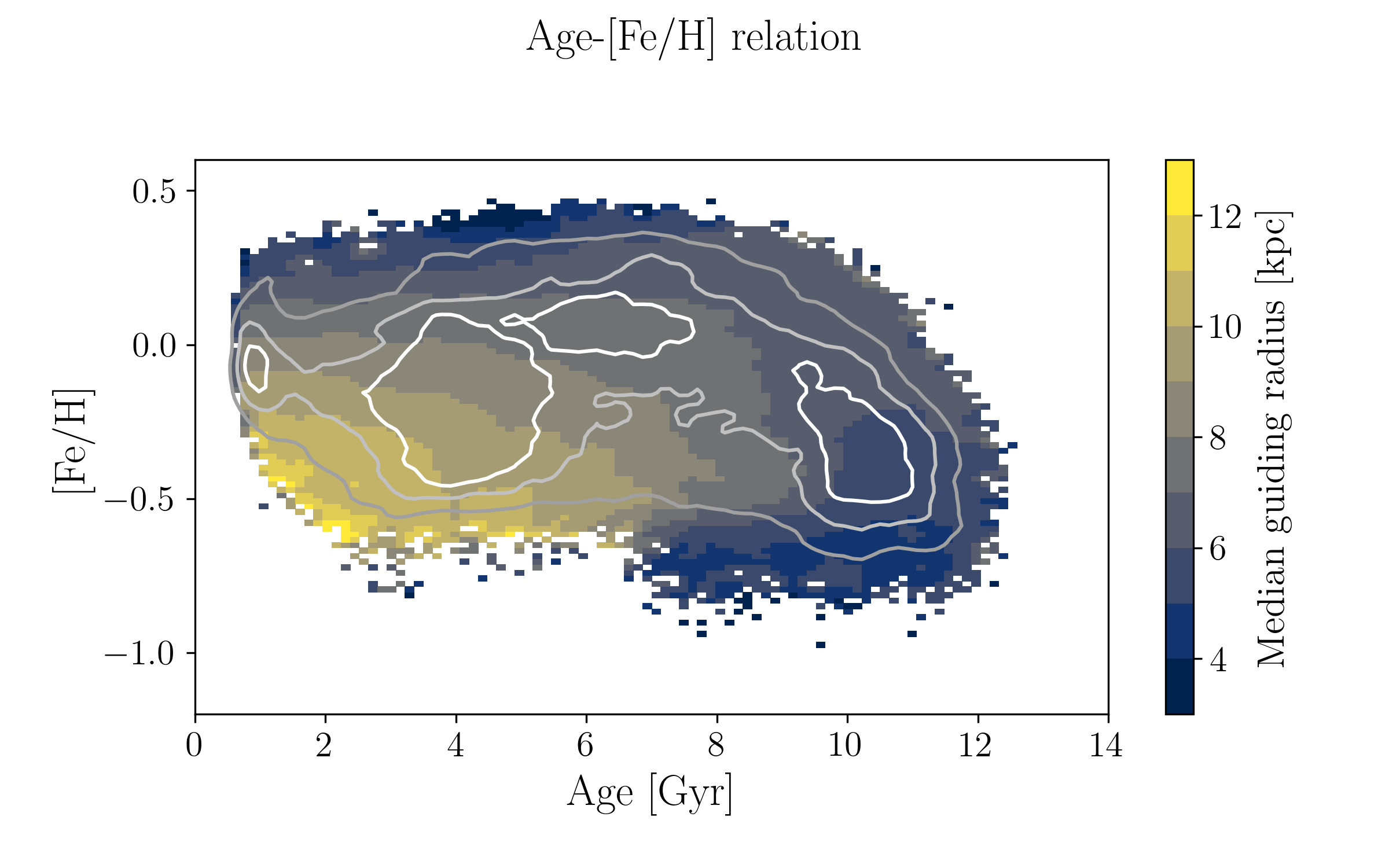}\\
\caption{\label{fig:agefe_R} 
Age-[Fe/H] relation, colour-coded by median guiding radius and smoothed with a Gaussian kernel. Overplotted are iso-density contours corresponding to 20, 40, and 60 stars per bin. Only bins containing at least three stars are shown.} 
\end{center}
\end{figure}

An important abundance ratio that was not used on purpose in our default {\tt XGBoost} model is [Fe/H]. We can therefore analyse the age-metallicity relation of our sample without too much fear of circularity (although one could plausibly argue that [Fe/H] is also directly related with the other abundance ratios; see e.g. \citealt{Ness2019}). 

Figure \ref{fig:agefe_R} shows the age-metallicity relation (AMR) of our whole APOGEE sample, colour-coded by median guiding-centre radius (estimated as $R_{\rm guide}=R_{\rm Gal}\cdot v_{\Phi} / v_{\rm circ})$, assuming a flat rotation curve ($v_{\rm circ}=229.76$ km/s; \citealt{Schonrich2010, Bovy2012_velocityCurve}) and using proper motion measurements from {\it Gaia} DR3 \citep{GaiaCollaboration2022Vallenari}. Density contours are overplotted. 
The overall picture is consistent with both predictions from classical multi-zone GCE models (e.g. \citealt{Chiappini2009}), GCE models with radial mixing (e.g. \citealt{Minchev2014, Kubryk2015, Johnson2021}), and cosmological zoom-in simulations of Milky-Way-like galaxies (e.g. \citealt{Renaud2021, Lu2022}): the colour-code clearly indicates a gradual inside-out formation of the disc within the last $\sim7$ Gyr, with a rapid enrichment in the inner disc (see also \citealt{Joyce2023}), and a rather recent onset of star formation in the outermost parts of the disc.

\begin{figure}
\begin{center} 
\includegraphics[width=.495\textwidth]{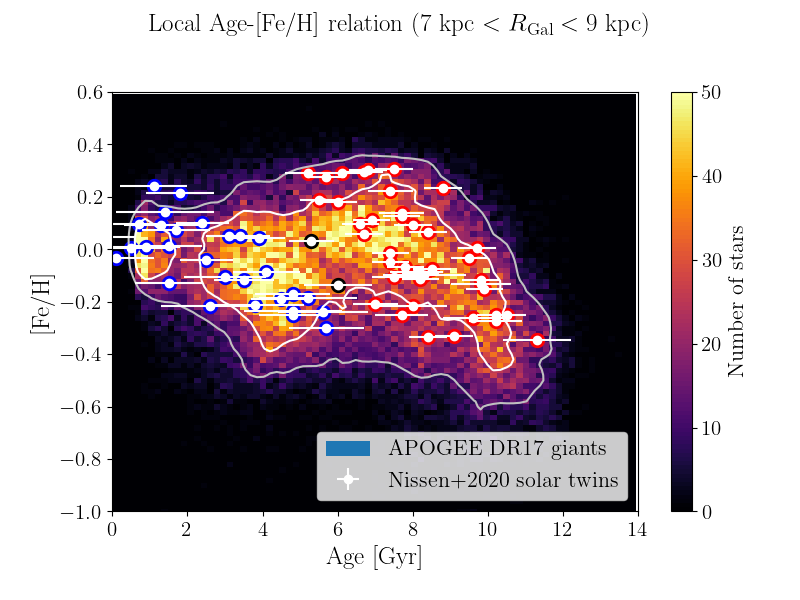}\\
\caption{\label{fig:agefe_local} 
Age-[Fe/H] relation for the solar vicinity (7 kpc $<R_{\rm Gal}<9$ kpc, $|Z_{\rm Gal}|<1$ kpc), as a 2D histogram and as smoothed iso-density contours corresponding to 10 and 30 stars per bin. Overplotted are the 72 solar twins studied by \citet{Nissen2020}, with the edge colour of the symbols corresponding to the groups defined by their Fig. 3.} 
\end{center}
\end{figure}

Restricting ourselves to the extended solar neighbourhood (7 kpc $<R_{\rm Gal}<9$ kpc, $|Z_{\rm Gal}|<1$ kpc), we obtain the local AMR shown in Fig. \ref{fig:agefe_local}. It resembles the one reconstructed from GALAH data by \citet[][see also \citealt{Queiroz2023}, Fig. 18]{Sahlholdt2022} and confirms the bimodality in the AMR hinted by \citet{Nissen2020} and later corroborated by the works of \citet{Jofre2021} and \citet{Xiang2022}. Following the terminology of the latter authors, the younger of the two branches in the AMR corresponds to the late, dynamically quiescent phase of disc evolution, while the older one comprises the in-situ halo (not present in our sample due to the [Fe/H]$>-1$ criterion) and the old, [$\alpha$/Fe]-enhanced disc. This is in line with traditional scenarios of dual disc formation (e.g. \citealt{Chiappini1997, Fuhrmann2017a}).
The fact that we clearly see the split in the AMR further validates the high precision of our age estimates.

However, we remind ourselves that the AMRs shown in Figs. \ref{fig:agefe_R} and \ref{fig:agefe_local} are a convolution of i. the underlying evolution of the ISM in the Milky Way disc over time, ii. stellar radial migration, iii. observational uncertainties (most importantly in terms of age), and iv. the selection function of our sample. Therefore, deciphering the past chemical-enrichment history is an endeavour that requires taking into account radial migration (e.g. \citealt{Minchev2018, Frankel2018, Frankel2019, Frankel2020, BeraldoeSilva2021, Lu2023, Ratcliffe2023}). 

\subsection{Spatial age trends}\label{sec:map}

\begin{figure}
\begin{center} 
\includegraphics[width=.495\textwidth]{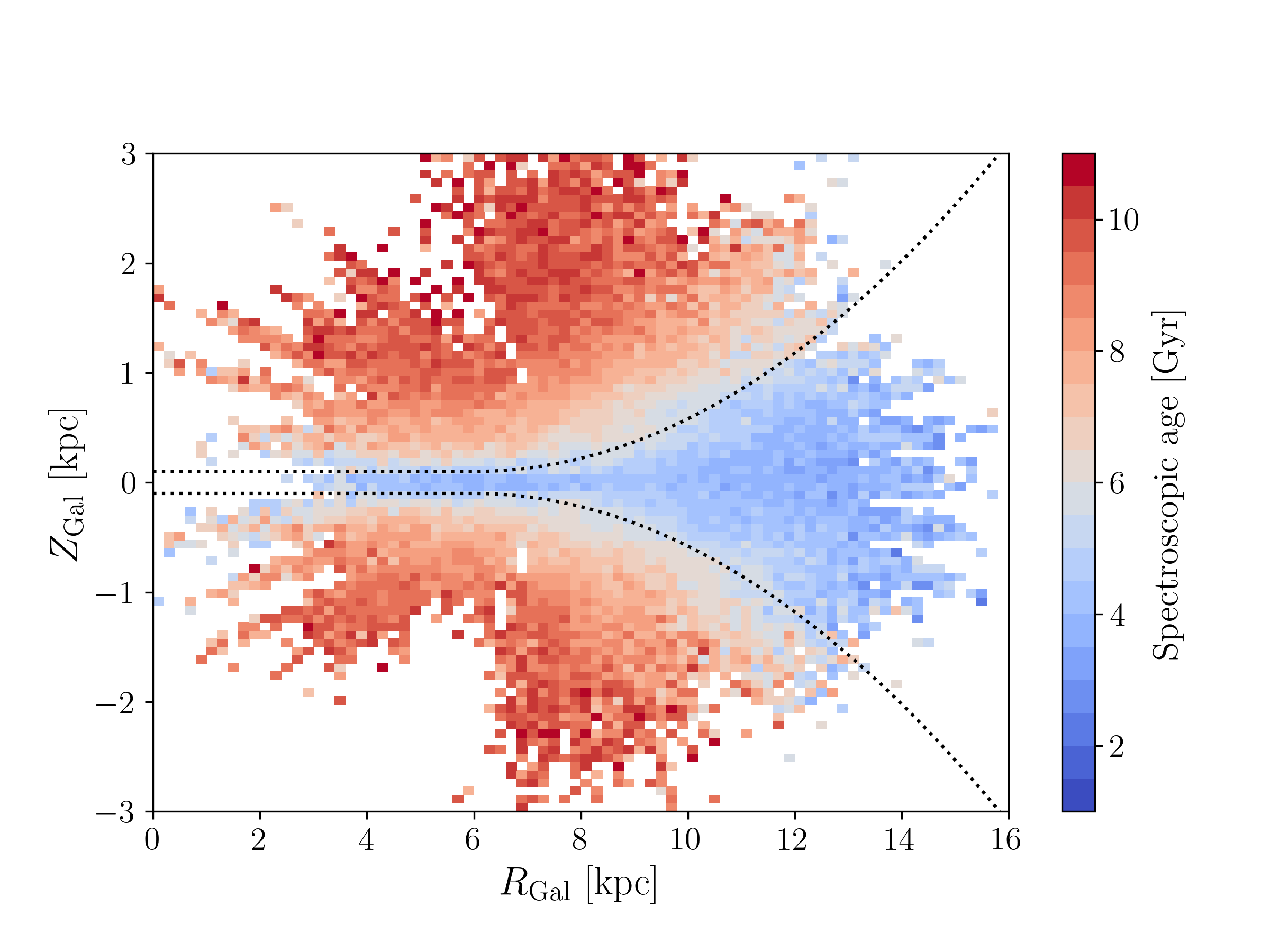}
\caption{\label{fig:rz_age} 
Median age as a function of position in Galactocentric coordinates. Only bins containing more than three stars are shown, no smoothing is applied. The dotted lines trace approximately the region in which the median age equals to 5 Gyr as a quadratic function of $R_{\rm Gal}$, clearly showing the effect of the young disc's flare. 
} 
\end{center}
\end{figure}

Another interesting plot is the median age map in cylindrical ($R_{\rm Gal}$ vs $Z_{\rm Gal}$) coordinates shown in Fig. \ref{fig:rz_age}. When interpreting this plot, we recall that in this work we have been extrapolating the Milky Way disc's intrinsic age-abundance relations from the {\it Kepler} field to the whole range of the Galaxy covered by our sample ($0\lesssim R_{\rm Gal}\lesssim16$ kpc). We also remember that selection effects have not been taken into account in the creation of this map (nor in any other of the shown plots). 

Even with these caveats, Fig. \ref{fig:rz_age} clearly shows the vertical age gradient we expect to see, and also the clear signature of the flared outer disc (as illustrated by the dotted lines that follow the 5 Gyr isochrone in this plot with a quadratic function starting at $R_{\rm Gal}\sim6$ kpc). In fact, not only the younger populations display flaring, but mono-age populations are expected to show nested flares, as argued by \citet{Minchev2015}. The prediction of their model was a strong age gradient in the morphological thick disk ($|Z_{\rm Gal}|>0.5$ kpc) that was indeed found in earlier APOGEE data (\citealt{Martig2016a}, see also \citealt{Imig2023}). We can now confirm this result with much larger statistics and extent in $R_{\rm Gal}$. This is in contrast to associating age with high-[$\alpha$/Fe] mono-abundance populations (MAPs), which show no flaring (\citealt{Bovy2016}, but see \citealt{Lian2022}) due to significant negative age gradients in a given MAP \citep{Minchev2017}.

\citet{Mackereth2017} studied the age-metallicity structure of the Milky Way disc using 31,000 APOGEE DR12 stars with [C/N] age estimates inferred by \citet{Martig2016}. The authors fitted parametric density profiles to the (volume-corrected) APOGEE star counts and found radially broken exponential radial profiles as well as flared exponential vertical profiles. The same behaviour could be reproduced by \citet{BeraldoeSilva2020} with a simulation including clumpy star formation in the early epochs of the disc. This big picture is still consistent with the APOGEE DR17 data (but see \citealt{Sysoliatina2022}).  

\begin{figure}
\begin{center} 
\includegraphics[width=.495\textwidth]{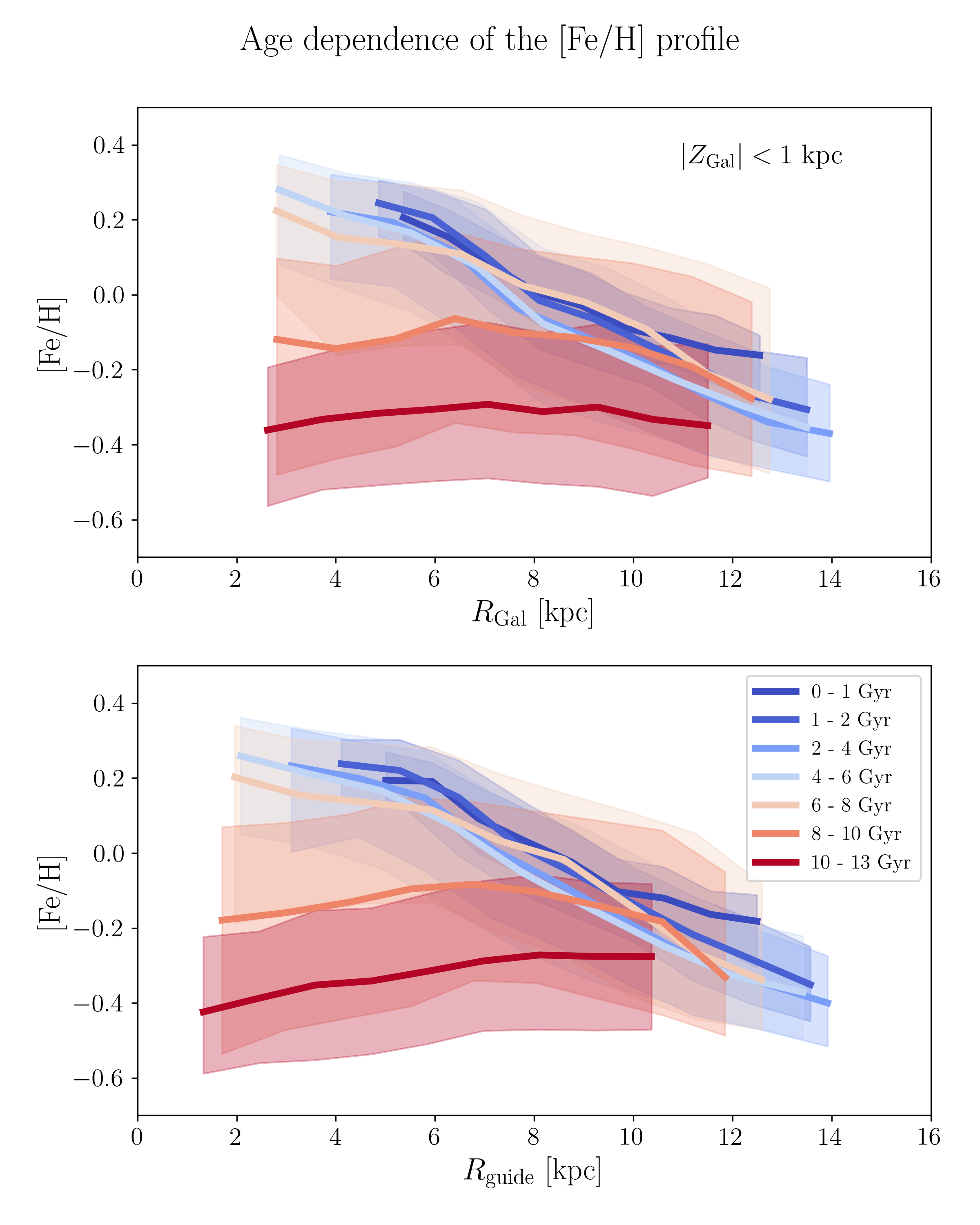}\\
\caption{\label{fig:rfe_age} 
Radial [Fe/H] profile in bins of age. Top panel: with respect to $R_{\rm Gal}$. Bottom panel: with respect to the guiding-centre radius, $R_{\rm guide}$. The youngest bin contains 2\,807 stars, all other bins more than 9\,000 stars.} 
\end{center}
\end{figure}

\subsection{Galactic radial [Fe/H] profile as a function of age}\label{sec:grad}

The Galactic radial abundance profile is an important observable that helps to constrain the chemical evolution of the Milky Way (e.g. \citealt{Trimble1975, Tinsley1980}). First discovered in nebular tracers \citep{Hawley1978, Peimbert1980, Afflerbach1997}, the presence of a negative radial abundance gradient in the Milky Way was confirmed in field stars soon thereafter \citep{Janes1979, Luck1980}. Since the 1980s, studies of the Galactic radial abundance profile as a function of age thrived, as larger samples of tracers with distance, age, and abundance were compiled. Most importantly, open clusters \citep[e.g.][]{Panagia1981, Lynga1982, Friel1995, Friel2002, Magrini2009} and planetary nebulae \citep[e.g.][]{Maciel2005, Stanghellini2010, Stanghellini2018} were used to infer the age dependence of the Milky Way's abundance profile, sometimes with contradictory results. With the advent of large spectroscopic surveys, it also became possible to use larger samples of field stars with isochrone ages to infer the age dependence of the abundance gradient \citep[e.g.][]{Nordstrom2004, Casagrande2011, Xiang2015}.

In parallel, also Galactic chemical-evolution modelling became more important and helped to explain the metallicity-gradient observations, typically by assuming an inside-out formation of the Galactic disc \citep[e.g.][]{Ferrini1994, Koeppen1994, Hou2000, Chiappini2001, Naab2006}. 
Nevertheless, the model comparison to different tracers can lead to opposite conclusions, since each tracer population is affected by different biases (in terms of age, distance, abundance, and selection effects). A more comprehensive review on the history of determinations of the age dependence of the radial metallicity gradient can be found in Sects. 1 and 6 of \citet{Anders2017a}.

\begin{figure}
\begin{center} 
\includegraphics[width=.495\textwidth]{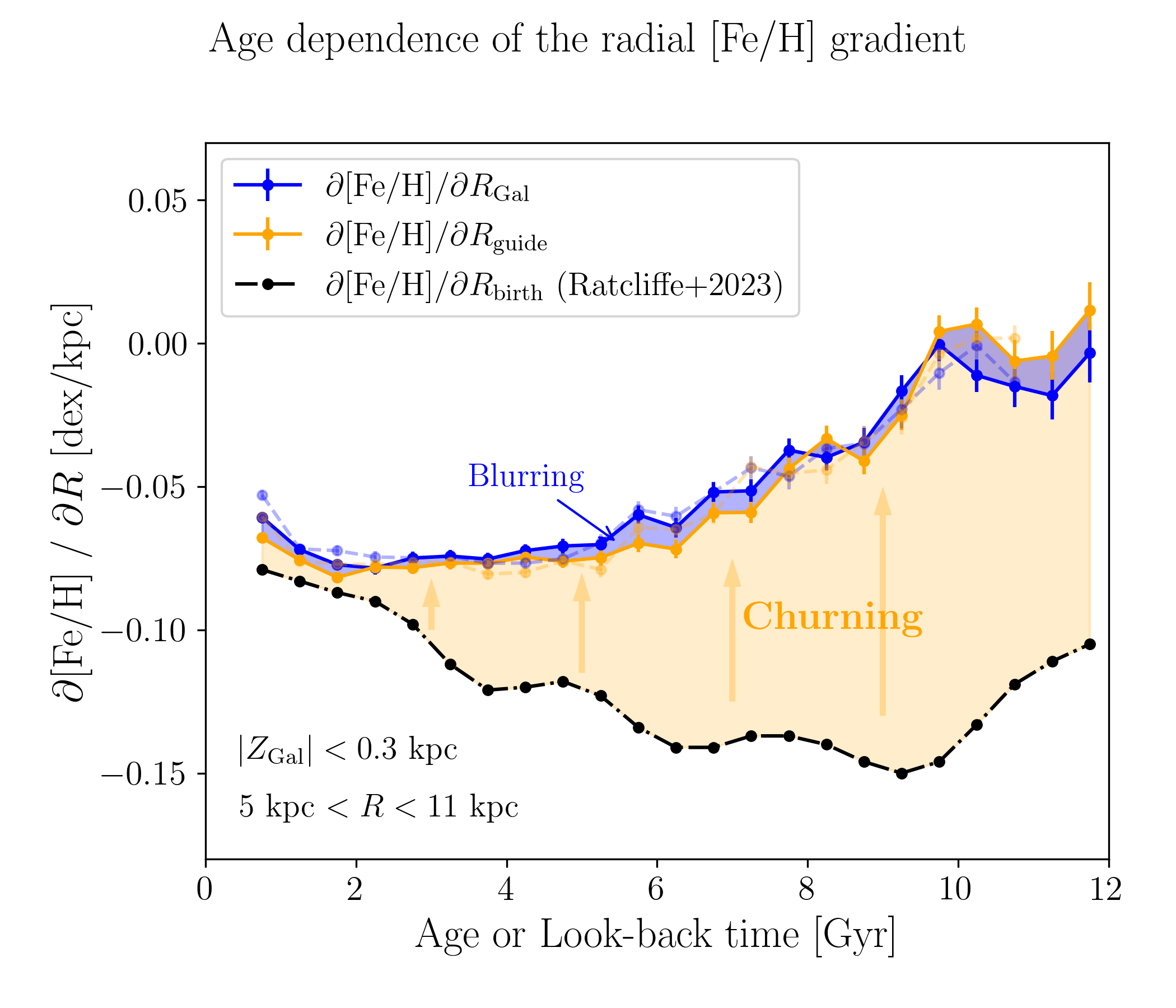}\\
\caption{\label{fig:dfedr_age} 
Age dependence of the radial [Fe/H] abundance gradient in terms of present-day Galactocentric distance (blue) and guiding-centre radii (orange). Each point was obtained by 3-parameter (slope, intercept, and dispersion) Bayesian fits to the [Fe/H]-$R$ distribution, using only data in the respective age bin, restricted to $|Z_{\rm Gal}|<0.3$ kpc and 5 kpc $<R_{\rm Gal / guide}<11$ kpc (39\,920 and 40\,629 stars, respectively). The faint dashed lines and errorbars correspond to the gradients when using the alternative set of ages obtained with {\tt XGBoost} quantile regression (see App. \ref{sec:datamodel}). In black we overplot the radial [Fe/H] gradients in the interstellar medium across look-back time obtained by \citet[][]{Ratcliffe2023}, calculated for the same age bins. The difference between the blue and the orange line (blue-shaded region) can be attributed to radial heating (blurring), while the difference between the orange and the black line (orange-shaded region) highlights the influence of radial migration (churning). Beyond $\sim 10$ Gyr, the gradients are consistent with zero.} 
\end{center}
\end{figure}

\citet{Anders2017a} studied the age dependence of the radial metallicity profile close to the Galactic plane ($|Z_{\rm Gal}|<0.3$ kpc) using a sample of 418 red-giant stars observed by both APOGEE (DR12; \citealt{Alam2015}) and the CoRoT satellite \citep{Baglin2006}. Thanks to our machine-learning approach we have now dramatically increased the sample size of field red giants with statistically meaningful ages to $>100,000$, obtaining a much less geometrically biased sample. We therefore reanalyse the age dependence of the Galactic radial metallicity profile in Figs. \ref{fig:rfe_age} and \ref{fig:dfedr_age}.

Figure \ref{fig:rfe_age} shows the radial [Fe/H] profile of the APOGEE red-giant sample in seven broad age bins. In  the top panel we display the [Fe/H] profiles (running median and 1$\sigma$ quantiles) as a function of current Galactocentric distance, while in the bottom panel we show the same as a function of guiding-centre radius. The differences between both panels, at least for the younger populations, are very subtle.

The radial metallicity distribution measured for the youngest populations ($0-1$ Gyr, $1-2$ Gyr, $2-4$ Gyr bins) are dominated by a strong linear ($\partial$[Fe/H]/$\partial R\simeq -0.06$ dex/kpc) gradient, with a typically symmetric dispersion that gradually increases with age. The gradient of the $1-5$ Gyr populations is also slightly steeper (by $0.015$ dex/kpc) than the gradient of the youngest bin ($<1$ Gyr), in agreement with previous results from field stars \citep[e.g.][]{Nordstrom2004, Casagrande2011, Anders2017a, Wang2019} and recent open-cluster studies \citep{Spina2021, Netopil2022, Myers2022, GaiaCollaboration2022Recio}, while in slight tension with other studies that found a rather monotonic flattening of the metallicity gradient with increasing age \citep{Chen2020, Vickers2021}. 

For older ages, it is expected that the radial metallicity gradient (and also the age-velocity dispersion relation, see Sect. \ref{sec:avr}) for open clusters behaves differently from field stars, due to a survival bias of clusters that migrate outwards \citep{Lynga1987, Anders2017a, Spina2021}.
Nevertheless, the importance of radial migration for the gradient evolution is still debated in the open-cluster community \citep{Spina2022, Magrini2023}. In the sample of field stars used in this work, we find that the [Fe/H] gradient significantly flattens (and even inverts when considering $R_{\rm guide}$) for the oldest age bins (Figure \ref{fig:rfe_age}). This weakening in the metallicity gradient, along with increased scatter about the running median, is an expected consequence of radial migration \citep{Kubryk2013, Minchev2013}, and is in agreement with other works \citep[e.g.][]{Casagrande2011, Xiang2015, Anders2017a}.

Unfortunately, the age dependence of the radial abundance gradient does not allow us to directly infer the evolution of the abundance gradient in the interstellar medium. In fact, just as in the case of the AMR discussed in Sect. \ref{sec:amr}, the abundance profiles shown in Fig. \ref{fig:rfe_age} are a convolution of Galactic chemical evolution, stellar mixing (heating and migration, or "blurring" and "churning" in the terminology of \citealt{Schonrich2009}), observational uncertainties, and selection effects.

\begin{figure}
\begin{center} 
\includegraphics[width=.495\textwidth]{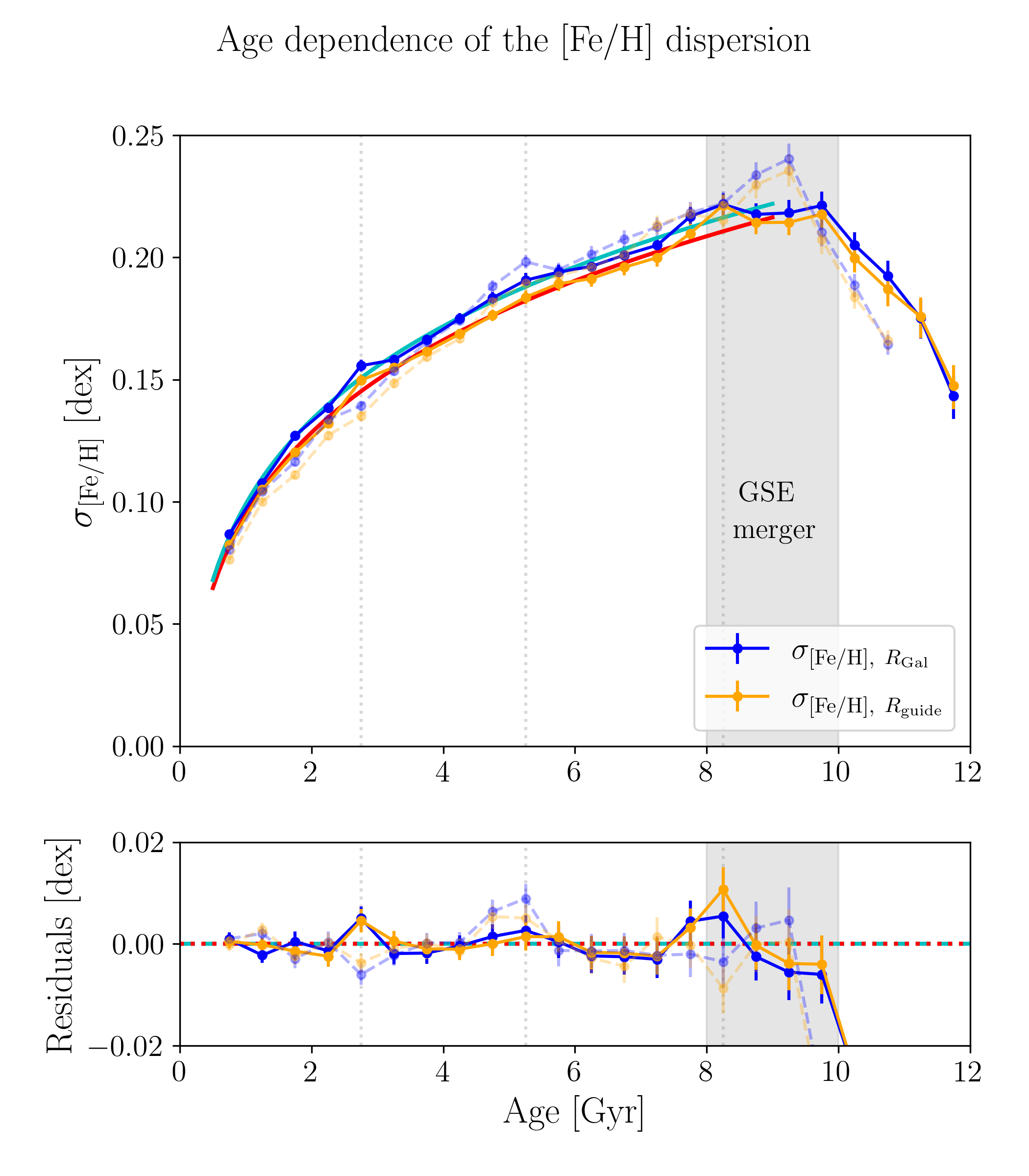}\\
\caption{\label{fig:sigfe_age} 
Age dependence of the [Fe/H] dispersion around the radial [Fe/H] abundance gradient shown in Fig. \ref{fig:dfedr_age}. Simple power-law fits to the data (for ages $<7$ Gyr) are overplotted, with power-law coefficients $0.12$ and $0.15$ for the dispersions about the present-day abundance gradient (cyan) and the abundance gradient with respect to the guiding radius (red), respectively. The lower panel shows the residuals with respect to these fits. As in Fig. \ref{fig:dfedr_age}, the faint dashed lines and errorbars correspond to the gradients when using the alternative set of ages obtained with {\tt XGBoost} quantile regression. The vertical lines highlight regions of departure from the power-law trend, possibly connected to episodes of enhanced star formation (or enhanced radial migration). The grey-shaded area marks the age interval in which we expect to see signatures from the Gaia Sausage Enceladus (GSE) merger event.} 
\end{center}
\end{figure}

Figure \ref{fig:dfedr_age} shows the result of Bayesian linear fits (see Appendix \ref{sec:appendix2} for details) to the radial [Fe/H] profiles in smaller age bins and restricted to a thinner slice of the Galactic disc ($|Z_{\rm Gal}|<0.3$ kpc) and an $R$ range between 5 kpc and 11 kpc. As in \citet{Anders2017a}, these fits account for both the observational uncertainties in [Fe/H] and an intrinsic abundance scatter about the linear trend. The present-day abundance gradients of mono-age populations in terms of Galactocentric distance ($\partial$[Fe/H]/$\partial R_{\rm Gal}$) are shown in blue, while the gradients in terms of guiding-centre radius ($\partial$[Fe/H]/$\partial R_{\rm guide}$) are shown in orange. For illustration, we also show the [Fe/H] gradient with respect to the birth radius inferred by \citet{Ratcliffe2023}, using the same age bins. These values can be thought of as a reconstruction of the Galactic radial abundance gradient in the interstellar medium (ISM) at a certain look-back time\footnote{The reconstruction of the radial abundance gradients as a function of look-back time is based on the method of \citet{Lu2023}. These authors found, using different suites of cosmological hydrodynamical simulations of Milky-Way-like galaxies, a tight anticorrelation between the ISM gradient at a certain look-back time and the $5\%-95\%$ interpercentile range of stellar metallicities found in the corresponding age interval.}. As indicated by the blue-shaded region in Fig. \ref{fig:dfedr_age}, the difference between the blue and the orange curves is produced by radial heating (blurring) of stellar orbits, flattening the observed radial abundance gradient only slightly in each age bin. On the other hand, the much larger difference between the black and the orange curve is produced by radial migration (churning), as highlighted by the orange-shaded area.

The picture drawn in Fig. \ref{fig:dfedr_age} depends slightly on the exact age scale used, as exemplified by the fits using the {\tt XGBoost} quantile regression ages (see Sect. \ref{sec:datamodel}) shown as faint dashed lines in that figure. It is also affected by the systematics discussed in Sect. \ref{sec:validation}, and in particular less trustworthy for older ages (both in terms of the age estimates and the estimated birth radii and abundance gradients; see e.g. \citealt{Lu2022a}). However, we see that radial migration is indeed a main driver in shaping the observed Galactic abundance gradient for mono-age populations, even for relatively small ages ($<1$ Gyr), and becoming more and more important for older ages (e.g. for ages older than 10 Gyr half of the solar neighbourhood's thin-disc stars should have migrated outward; \citealt{BeraldoeSilva2021}. This picture is also in excellent agreement with the main result of \citet{Frankel2020} who concluded that the secular orbit evolution in the Galactic disc is dominated by diffusion in angular momentum, with radial heating being less important by one order of magnitude. A deeper analysis of the actual evolution of the Galactic abundance gradients (not only for [Fe/H] but also for other elements) using our age estimates is presented in \citet{Ratcliffe2023}, revealing substructure in the evolution of the abundance gradients that possibly correspond to enhanced star formation episodes triggered by passages of the Sgr dSph satellite and the GSE (see the wiggles in the black curve in Fig. \ref{fig:dfedr_age}).

Our Bayesian fits of the radial metallicity gradient between 5 and 11 kpc in age bins of 0.5 Gyr (shown in Fig. \ref{fig:dfedr_age}) also deliver precise estimates of the [Fe/H] dispersion about the gradient. Figure \ref{fig:sigfe_age} shows these dispersion measurements as a function of age, again for both cases, with respect to $R_{\rm Gal}$ and $R_{\rm guide}$. As expected (and first demonstrated by \citealt{Spina2021} for open clusters), the dispersions about the radial abundance gradient with respect to the guiding radius are consistently smaller by a small amount (typically $0.01-0.02$ dex). 

In addition, thanks to the dramatically increased sample size and therefore much smaller statistical uncertainties, we can now quantify the behaviour of the [Fe/H] dispersion over time much better than in \citet{Anders2017a}. We find that the increase of the abundance dispersion between 1 and 7 Gyr can be well fit by a power law with an exponent around $\sim0.15$ (solid coloured lines in Fig. \ref{fig:sigfe_age}). For even younger ages ($<500$ Myr), this relation is expected to flatten, since the intrinsic [Fe/H] dispersion of younger tracers is finite and not much smaller than 0.05 dex \citep{Genovali2014, Wenger2019, Spina2021}. However, our red-giant sample does not cover this age range. The increase in abundance scatter with age is obviously created by radial migration \citep[e.g.][]{Haywood2008, Schonrich2009, Minchev2014, Kubryk2015}. The smooth functional form of this increase with age suggests that radial migration (churning) can be described by a sub-diffussive process with a constant efficiency over a large phase of the Galactic disc's lifetime \citep[see also][]{Anders2020}. 

Hints of departures from the power-law trends in Fig. \ref{fig:sigfe_age} are seen in the age bins 2.5--3 Gyr, 5--5.5 Gyr, and possibly 8--8.5 Gyr. For example, an enhanced [Fe/H] dispersion with respect to the overall trend can be appreciated around 2.5--3 Gyr (when using our fiducial age scale). Interestingly, this age corresponds to an epoch for which an enhanced star formation rate has been reported by several studies using {\it Gaia} \citep{Bernard2018, Mor2019, Isern2019, Alzate2021} and spectroscopic survey data \citep{Johnson2021, Sahlholdt2022, Spitoni2023}. Another small enhancement in $\sigma_{{\rm [Fe/H]},\ R_{\rm Gal}}$ and $\sigma_{{\rm [Fe/H]},\ R_{\rm guide}}$ seems to occur in the 5--5.5 Gyr bin (in our age scale), coinciding with an epoch of enhanced star formation proposed in the literature \citep{Ruiz-Lara2020, Alzate2021, Sahlholdt2022}, possibly related to the first passage of the Sgr dSph galaxy \citep{Ruiz-Lara2020}. Recent simulations indeed suggest that Sgr passages can temporally enhance migration \citep{Carr2022}. As also demonstrated in Fig. \ref{fig:sigfe_age}, however, these small departures from the main power-law trend are not always robust to a change in the used age scale. 


The grey-shaded area in Fig. \ref{fig:sigfe_age} marks the approximate time frame in which the last significant merger event, the {\it Gaia} Sausage Enceladus (GSE) merger \citep{Helmi2018, Belokurov2018}, probably affected the Milky Way disc. For example, \citealt{Chaplin2020} found an upper limit of $>11.0\pm1.5$ Gyr by age-dating the in-situ halo star $\nu$ Ind. \citet{Montalban2021} tightened this constraint further using detailed asteroseismic modelling of $96$ old stars observed with {\it Kepler} and found that the merger happened $\sim 10$ Gyr ago. 

In the oldest regime ($\gtrsim9-10$ Gyr in our age scale), we see a sharp decrease of $\sigma_{\rm [Fe/H]}$ with age (i.e. an increase with cosmic time). This could potentially be related to the turbulent formation of the early Milky Way disc, either through numerous mergers \citep[e.g.][]{Brook2004, Brook2005} or by internal interactions in turbulent and clumpy discs \citep[e.g.][]{Bournaud2009, Forbes2012, Amarante2020, Bird2021}, or both \citep{Belokurov2022}. 
However, the expectation from recent Milky-Way-like simulations such as FIRE-2 \citep{Wetzel2023}, NIHAO-UCD \citep{Buck2020}, or Auriga \citep{Grand2017} is rather that rapid variations in star formation, outflow rate, and depletion time during early stages of evolution result in large variations in the elemental abundances of old stars, especially for [Fe/H]$<-1$ \citep[][Fig. 19 - although their analysis is not strictly focused on the disc and concerns the total metallicity spread, not subtracting the gradient]{Belokurov2022}. Another option is that the GSE merger triggered a burst in star formation in the Galactic disc \citep{An2022, Ciuca2022}, as a result of which we see an enhanced [Fe/H] dispersion about the radial abundance gradient of the $\sim 8-9$ Gyr red-giant population that then drops gradually. We note, however, that the enhanced [Fe/H] dispersion can also be produced by a temporally enhanced radial migration efficiency (again related to the merger). We also recall that in the old age regime our disc sample can potentially be biased by the metallicity range for which our age estimation is valid ([Fe/H]$>-1$). Again, a full forward modelling (and/or a training set including enough metal-poor stars) is necessary to reliably extend this analysis into the oldest regime.

\subsection{Age-velocity dispersion relation}\label{sec:avr}

\begin{figure}
\begin{center} 
\includegraphics[width=.495\textwidth]{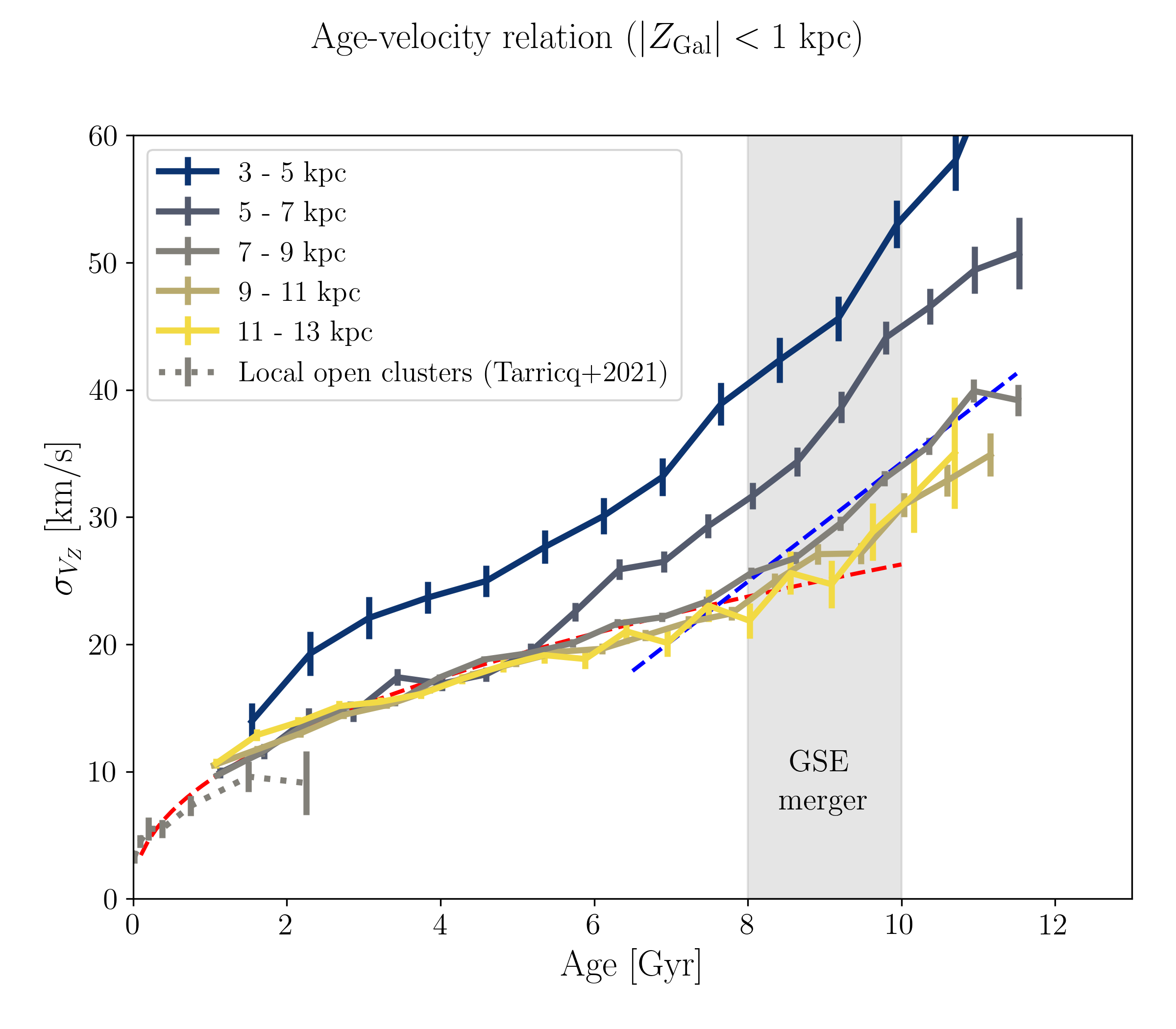}\\
\caption{\label{fig:sigvz_age} 
Age - velocity dispersion relationship (AVR), in bins of Galactocentric distance, for stars close to the Galactic plane ($|Z_{\rm Gal}|<1$ kpc). Also plotted are the results of \citet{Tarricq2021} for open clusters in the solar vicinity. The red dashed line corresponds to a simple power-law fit ($\beta=0.46$) for ages $<7$ Gyr in the Galactocentric distance bin 7-9 kpc, while the blue dashed line corresponds to a linear fit for ages between 7 and 11 Gyr in the same $R_{\rm Gal}$ bin. The shaded region highlights the age range in which we see a steepening in the AVR, potentially related to the GSE merger event. }
\end{center}
\end{figure}

While our age estimates critically depend on the quality of the APOGEE stellar parameters and abundances (except for [Fe/H]), they are completely independent from kinematics and position in the Galaxy. As an example for a kinematic relationship with age that is often used in Galactic archaeology, we show in Fig. \ref{fig:sigvz_age} the dispersion in vertical velocity $v_Z$ as a function of age (here restricted to $|Z_{\rm Gal}|<1$ kpc) -- often called the age-velocity relation (AVR).

Since \citet{Gliese1956} and \citet{vonHoerner1960} found significant differences in the motions of nearby stars as a function of their spectral type, the AVR in the solar neighbourhood is an important observable that constrains the vertical heating history of the Milky Way's disc. It has been studied by many authors using a variety of tracers in the past 50 years (e.g. \citealt{Wielen1974, Mayor1974, Byl1974, Meusinger1991, Nordstrom2004, Casagrande2011, Aumer2016, Ting2019, Raddi2022}).

Figure \ref{fig:sigvz_age} shows the AVR of our APOGEE DR17 sample in 2 kpc wide bins of Galactocentric distance between 3 kpc and 13 kpc, with each bin containing more than 14\,000 stars (except for the 3-5 kpc bin with 6\,000 stars and a rather distinct AVR behaviour, probably due to the influence of the Galactic bar; \citealt{Saha2010, Grand2016}). Since the age range of our sample does not cover the youngest stars, we also plot the AVR recently obtained by \citet{Tarricq2021} using a sample of 418 {\it Gaia}-confirmed OCs with radial velocities in the solar neighbourhood. The authors found that the AVR in the young regime can be described by a power law with $\beta\approx0.2$. Although the OC sample has thus much lower statistics and is possibly affected by the cluster destruction bias in the 2--3 Gyr bin \citep{Anders2017a, Spina2022}, we see that the trend of growing velocity dispersion continues in our field star sample, regardless of Galactocentric distance. Indeed, we find that the trend in our age range between 0.5 Gyr and 7 Gyr can be well described with a power law for $R_{\rm Gal}>5$ kpc (in accordance with e.g. \citealt{Aumer2016, Ting2019} or the simulation of \citealt{Agertz2021}), but with a notably different power-law index from the OC sample ($\beta\approx 0.46$ for the 7-9 kpc bin; red dashed line in Fig. \ref{fig:sigvz_age}).

In part, the higher velocity dispersion for older stars in the inner disc seen in Fig. \ref{fig:sigvz_age} is expected from the exponential disc profile in galactic discs (for the case of a flat rotation curve and hydrostatic equilibrium: $\sigma_{v_{z}}(R) \propto \sqrt{\Sigma(R)\cdot h_z}$; e.g. Eq. 7 in \citealt{vanderKruit2011}). We also see hints (in the smallest age bin) of the expected trend that stars in the outer disc are born with a higher velocity dispersion (due to the smaller vertical force acting on the gas in the outer disc). However, the velocity dispersions in the intermediate age regime ($\sim2-6$ Gyr) are surprisingly congruent from 5 kpc out to 13 kpc (different from what is seen in modern simulations; e.g. Fig. 14 in \citealt{Agertz2021}), which possibly indicates that vertical heating happened with similar efficiency over a large part of the Galactic disc within the past $\sim 6-8$ Gyr. We caution, however, that age errors will always tend to smooth out any abrupt changes in the AVR (e.g. \citealt{Martig2014}).

For $R_{\rm Gal}>7$ kpc populations older than $\sim8$ Gyr, we see evidence for a departure from the power-law behaviour observed between 0.5 and 7 Gyr, in the sense that the AVR steepens (in the 5-7 kpc bin, a first steepening happens already around 5.5 Gyr in our age scale). This change of trend is exemplified by the blue dashed line in Fig. \ref{fig:sigvz_age}, which corresponds to a linear fit of the AVR in the 7-9 kpc bin restricted to the age range between 7 and 11 Gyr. Similar broken trends in the AVR can be seen in all $R_{\rm Gal}$ bins. Except for the innermost $R_{\rm Gal}$ bin, the exact onset of the steepening appears to depend slightly on $R_{\rm Gal}$. Similar trends for the solar vicinity have already been observed previously (e.g. \citealt{Meusinger1991, Quillen2000, Yu2018a}. Indeed, \citet{Quillen2000} already found, using the data of \citet{Edvardsson1993} together with the revised age estimates from \citet{Ng1998} an abrupt increase by a factor of two in the stellar velocity dispersions at an age of $9\pm1$ Gyr and proposed that the Milky Way suffered a minor merger 9 Gyr ago, which created the thick disc. This means that we have not discovered a new feature of the AVR, but confirmed a previously observed trend with much better statistics and individual age precision. In addition, we can now study the AVR over a much larger portion of the Galactic disc, finding tendencies with $R_{\rm Gal}$ that can potentially be used for a more detailed comparison to chemo-dynamical models of the Milky Way.

On the simulation side, the AVR has been shown to be affected by both internal and external kinematic processes. For example, \citet{Grand2016} found, using cosmological zoom-in simulations of Milky Way-sized galaxies, that in most cases the Galactic bar is the dominant heating agent, while spiral arms, radial migration, and adiabatic heating play a secondary role. In some of the \citet{Grand2016} simulations, however, the strongest source of vertical heating were external perturbations from massive satellites. Using cosmological simulations, \citet{Martig2014} showed that the AVR depends on the merger history at low redshift, and that jumps in the solar vicinity's AVR often correspond to (minor) merger events. However, \citet{Martig2014} also showed that in the presence of age errors these jumps are considerably blurred, so that only the largest merger events are detectable (and, depending on the age error, not as jumps but as changes in slope of the AVR). \citet{Yurin2015} also found that substructures are important for heating the outer parts of stellar discs but do not significantly affect their inner parts. \citet{Saha2010} related the vertical heating exponent in the inner parts of galaxies to the growth rate of the bar potential, while in the outer regions they found that vertical heating was often dominated by transient spiral waves and mild bending waves. 

Regarding the break/steepening in the AVR that we see in Fig. \ref{fig:sigvz_age} around $\sim 8$ Gyr, some work relying on simulations suggests that this is an effect of a GSE-like merger event \citep{Quillen2000, Martig2014}, while others associate it with the gradual settling of the early, turbulent Galactic disc \citep{Bird2013, Navarro2018, Bird2021, Yu2021}. We suggest that this controversy could potentially be settled with the data presented in this work (that cover an extensive range of the Galactic disc), by comparing the AVR in radial bins to simulations, taking into account selection effects.

The kinematics of mono-age, mono-metallicity APOGEE populations were studied in more detail by \citet{Mackereth2019}, albeit with less data (65\,719 stars) and not showing a prominent steepening in the AVR. Their results depend in part on the accuracy of the inferred ages (obtained using a Bayesian convolutional neural network and APOGEE DR14 spectra). We note, however, that the AVR, as other relationships with age, is not only blurred by statistical age errors, but also affected by systematic age errors and radial migration. Its full meaning is more clearly revealed when separating stars by their birth radii \citep{Minchev2018, Bird2021}. We also note that a more robust analytical description of the vertical heating history of the Galactic disc can be obtained by analysing the stellar kinematics in action space \citep{Ting2019}.

\section{Conclusions}\label{sec:conclusions}

In this paper we successfully use the {\tt XGBoost} algorithm to accurately estimate spectroscopic ages for 178\,825 APOGEE DR17 red-giant stars located close to the red clump ($2.2<\log g<3.4$, 4400 K$ <T_{\rm eff}<5200$ K) with a median statistical uncertainty of 17\%. These estimates are formally independent of [Fe/H]. In accordance with previous studies, we find that our red-giant age estimates are mostly driven by $T_{\rm eff}$, $\log g$, [C/Fe], and [N/Fe]. To verify the obtained ages (which are tied to the asteroseismic age scale of \citealt{Miglio2021}), we compare our results with the age scale of open clusters, asteroseismic+spectroscopic red-giant samples observed with CoRoT, K2, and TESS, as well as other machine-learning age estimates from the literature, finding acceptable agreement, modulo some (expected) systematic trends, with most of the test samples (none of which can be regarded as absolute benchmarks). Our analysis is reproducible via github\footnote{\url{https://github.com/fjaellet/xgboost_chem_ages}} and the age catalogue (see Appendix \ref{sec:datamodel}) can be accessed in that repository\footnote{\url{https://github.com/fjaellet/xgboost_chem_ages/blob/main/data/spec_ages_published.fits}} or via CDS. 

We then investigate some chemo-kinematic relationships with stellar age that are often used in Galactic archaeology. We find that the ages are precise enough to confirm the split in the local age-metallicity relation found by \citet{Nissen2020} using high-resolution spectroscopy of solar analogues (Fig. \ref{fig:agefe_local}). Analysing the Milky Way disc's age map (Fig. \ref{fig:rz_age}), we find a clear imprint of the flaring in the outer disc, impressively confirming that our age-estimation method (which is completely independent of sky position and distance) works well for the full range probed by the APOGEE DR17 giant population. 

In Sect. \ref{sec:grad} (Figs. \ref{fig:rfe_age}-\ref{fig:sigfe_age}), we analyse the Milky Way disc's radial metallicity profile. We present new and precise measurements of the Galactic radial metallicity gradient close to the Galactic plane ($|Z_{\rm Gal}|<0.3$ kpc, 5 kpc $<R_{\rm Gal}<11$ kpc) in 0.5 Gyr bins between 0.5 and 12 Gyr, confirming a steeper metallicity gradient for 2-5 Gyr old populations and a subsequent flattening for older populations mostly produced by radial migration (churning). 
In addition, our fits to the radial [Fe/H] profile allow us to analyse, with unprecedented precision, the dispersion about the radial abundance gradient as a function of age (Fig. \ref{fig:sigfe_age}). We see a clear power-law trend (with a power-law index $\beta\approx0.15$), indicating a smooth radial migration history in the Galactic disc over the past $\sim7$ Gyr. Departures from this power law are detected at ages of $\simeq9$ Gyr (possibly related to the Gaia Sausage Enceladus merger) and $\simeq 5$ Gyr (with lower significance, possibly related to a passage of the Sgr dSph galaxy; \citealt{Carr2022}). 

Finally, as an example of a kinematic relationship with age, we study the age-velocity dispersion relation (AVR) in 2 kpc wide radial bins around the Galactic centre (Fig. \ref{fig:sigvz_age}). We find that the AVR for ages up to $\sim7$ Gyr can be well described by a power law with an exponent around 0.5, and confirm earlier measurements reporting a pronounced steepening of the AVR at around $\sim9$ Gyr. Indeed, we find that this behaviour extends over a large extent of the Galactic disc (5 kpc $<R_{\rm Gal}<13$ kpc), while in the inner disc ($R_{\rm Gal}<5$ kpc) the AVR shows a more complex behaviour that we attribute to efficient heating by the Galactic bar. 

Our results reproduce the expected chemical, positional, and kinematic trends with age. This suggests that chemical clocks, and more broadly speaking, weak chemical tagging, is a viable method in Galactic Archaeology. The use of precision asteroseismology coupled with efficient machine-learning algorithms such as {\tt XGBoost} may allow for greater efficiency and accuracy in estimating ages for millions of stars in the era of large spectroscopic surveys.
We recall that some of our conclusions (for example, the exact time dependence of the radial abundance gradient) are sensitive to our absolute age scale and the systematics it may be suffering from (see Section \ref{sec:validation}).
The knowledge of precise and accurate ages, kinematics, and detailed chemical abundances of stars still holds the key to uncovering the formation and evolution of our Galaxy.

\begin{acknowledgements}
This work was partially funded by the Spanish MICIN/AEI/10.13039/501100011033 and by the "ERDF A way of making Europe" funds by the European Union through grant RTI2018-095076-B-C21 and PID2021-122842OB-C21, and the Institute of Cosmos Sciences University of Barcelona (ICCUB, Unidad de Excelencia ’Mar\'{\i}a de Maeztu’) through grant CEX2019-000918-M. FA acknowledges financial support from MCIN/AEI/10.13039/501100011033 through grants IJC2019-04862-I and RYC2021-031638-I (the latter co-funded by European Union NextGenerationEU/PRTR). B.R. and I.M. acknowledge support by the Deutsche Forschungsgemeinschaft under the grant MI 2009/2-1. CC acknowledges support from the Jesús Serra foundation. J.A.S.A. acknowledges funding from the European Research Council (ERC) under the European Union’s Horizon 2020 research and innovation programme (grant agreement No. 852839). H.P. acknowledges support from FAPESP proc. 2018/21250-9 and 2022/04079-0.

This project was developed in part at the Lorentz Center workshop Mapping the Milky Way, held 6-10 February, 2023 in Leiden, the Netherlands. We thank Ioana Ciuc{\u{a}} (Canberra) for sharing her APOGEE DR17 BINGO results, and Letizia Stanghellini (Tucson) and Tiancheng Sun (Beijing) for sending comments on the first version of the manuscript. Finally, we thank the anonymous referee for a very useful report that helped improve the quality of the paper.

The preparation of this work has made use of TOPCAT \citep{Taylor2005}, NASA's Astrophysics Data System Bibliographic Services, as well as the open-source Python packages \texttt{xgboost} \citep{Chen2016}, \texttt{astropy} \citep{Astropy2018}, and \texttt{NumPy} \citep{VanderWalt2011}. The figures in this paper were produced with \texttt{matplotlib} \citep{Hunter2007}. 

This work has made use of data from the European Space Agency (ESA) mission \textit{Gaia} (www.cosmos.esa.int/gaia), processed by the \textit{Gaia} Data Processing and Analysis Consortium (DPAC, www.cosmos.esa.int/web/gaia/dpac/consortium). Funding for the DPAC has been provided by national institutions, in particular the institutions participating in the \textit{Gaia} Multilateral Agreement. 

Funding for the Sloan Digital Sky Survey IV has been provided by the Alfred P. Sloan Foundation, the U.S. Department of Energy Office of Science, and the Participating Institutions. SDSS-IV acknowledges support and resources from the Center for High-Performance Computing at the University of Utah. The SDSS web site is \url{www.sdss.org}.

SDSS-IV is managed by the Astrophysical Research Consortium for the Participating Institutions of the SDSS Collaboration including the Brazilian Participation Group, the Carnegie Institution for Science, Carnegie Mellon University, the Chilean Participation Group, the French Participation Group, Harvard-Smithsonian Center for Astrophysics, Instituto de Astrof\'isica de Canarias, The Johns Hopkins University, 
Kavli Institute for the Physics and Mathematics of the Universe (IPMU) / University of Tokyo, Lawrence Berkeley National Laboratory, 
Leibniz-Institut f\"ur Astrophysik Potsdam (AIP),  
Max-Planck-Institut f\"ur Astronomie (MPIA Heidelberg), 
Max-Planck-Institut f\"ur Astrophysik (MPA Garching), 
Max-Planck-Institut f\"ur Extraterrestrische Physik (MPE), 
National Astronomical Observatory of China, New Mexico State University, 
New York University, University of Notre Dame, 
Observat\'ario Nacional / MCTI, The Ohio State University, 
Pennsylvania State University, Shanghai Astronomical Observatory, 
United Kingdom Participation Group,
Universidad Nacional Aut\'onoma de M\'exico, University of Arizona, 
University of Colorado Boulder, University of Oxford, University of Portsmouth, 
University of Utah, University of Virginia, University of Washington, University of Wisconsin, Vanderbilt University, and Yale University.

\end{acknowledgements}

\bibliographystyle{aa} 
\bibliography{chemical_ages}

\begin{appendix}

\section{Age catalogue}\label{sec:datamodel}

Table \ref{datamodel} shows the datamodel for the catalogue that we release with this work. Columns 2-4 correspond to the fiducial results obtained with the default {\tt XGBoost} (package version 1.7.6) run described in Sect. \ref{sec:method}. Columns 6-9 correspond to results obtained with {\tt XGBoost} quantile regression\footnote{\url{https://scikit-learn.org/stable/auto_examples/
ensemble/plot_gradient_boosting_quantile.html}}, available in package version 2.0.0-dev. The advantage of these latter results is that they come with more meaningful individual uncertainties, the disadvantage is that the median values suffer from a greater regression-towards-the-mean effect, i.e. noticeably smaller ages in in the old regime. It also results in a higher number of young $\alpha$-rich stars (1\,222 vs. 706). For most other stars, the differences between the two methods are small. In most of this paper, as well as in \citet{Ratcliffe2023}, we mainly use the columns {\tt spec\_age\_xgb} and {\tt spec\_age\_xgb\_uncert}.

Some of the caveats described in Sect. \ref{sec:caveats} are encoded as warning flags in the human-readable column {\tt spec\_age\_xgb\_flag} ({\tt spec\_age\_xgb\_quantilereg\_flag} for the quantile regression ages). For the 5\,383 affected stars (5\,549 in the quantile regression case) it can take the values {\tt BLUER\_THAN\_TRAINING\_SET}, {\tt REDDER\_THAN\_TRAINING\_SET}, {\tt HIGH\_VSINI}, {\tt APPARENTLY\_YOUNG\_ALPHA\_RICH}, and combinations thereof. The preparation steps necessary for the production of the age catalogue can be reproduced and modified following the procedure in the accompanying jupyter notebooks.

\begin{footnotesize}
\begin{table*}
\begin{center}
\caption{Data model of our APOGEE DR17 spectroscopic age catalogue.}
\begin{tabular}{llll}
\# & Column name &  Format & Description \\
\hline
1 & {\tt APOGEE\_ID}      & String (19A) & APOGEE ID (DR17) \\
2 & {\tt spec\_age\_xgb} & Float        & Spectroscopic age from fiducial {\tt XGBoost} model \\
3 & {\tt spec\_age\_xgb\_calib} & Float        & Calibrated spectroscopic age (see Fig. \ref{fig:uncerts}, top panel) \\
4 & {\tt spec\_age\_xgb\_uncert} & Float       & Age uncertainty estimate (see Fig. \ref{fig:uncerts}, bottom panel) \\
5 & {\tt spec\_age\_xgb\_flag} & String (51A) & Human-readable warning flag for potentially problematic stars\\ 
6 & {\tt spec\_age\_xgb\_quantilereg} & Float        & Spectroscopic age from {\tt XGBoost} quantile regression \\
7 & {\tt spec\_age\_xgb\_quantilereg\_calib} & Float        & Calibrated age from XGBoost quantile regression \\
8 & {\tt spec\_age\_xgb\_quantilereg\_sigl} & Float        & Lower 1$\sigma$ uncertainty from XGBoost quantile regression \\
9 & {\tt spec\_age\_xgb\_quantilereg\_sigu} & Float        & Upper 1$\sigma$ uncertainty from XGBoost quantile regression \\
10 & {\tt spec\_age\_xgb\_quantilereg\_flag} & String (52A) & Human-readable warning flag for potentially problematic stars\\ 
\hline
\end{tabular}
\label{datamodel}
\end{center}
\end{table*}
\end{footnotesize}

\section{Additional age maps}

\begin{figure*}
\begin{center} 
\includegraphics[width=.95\textwidth]{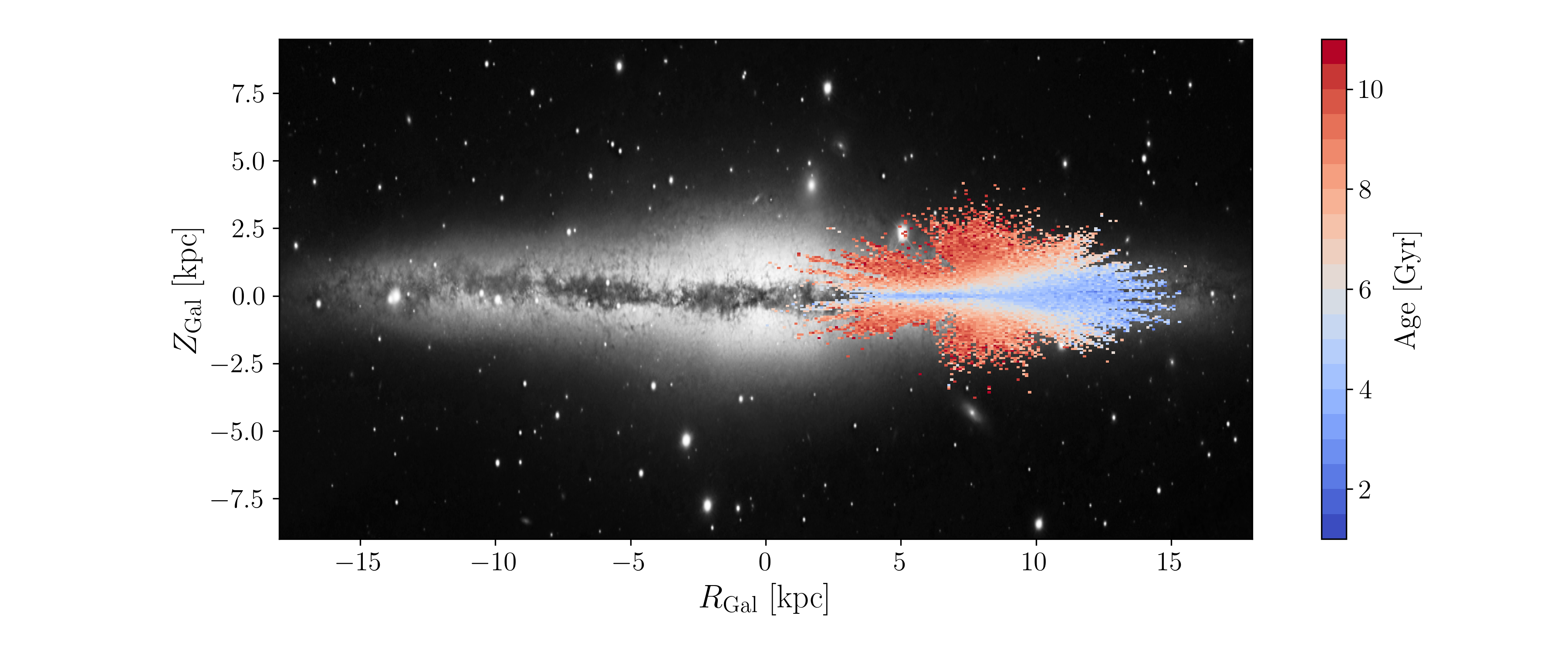}\\
\caption{\label{fig:rz_age2} 
Median age as a function of position in Galactocentric coordinates, similar to Fig. \ref{fig:rz_age}, but using equal scales on both axes and showing a greater extent of the Galactic disc. We also show an $r$-band image of the nearby edge-on galaxy NGC 891 in the background for illustration.
} 
\end{center}
\end{figure*}

\begin{figure*}
\begin{center} 
\includegraphics[width=.8\textwidth]{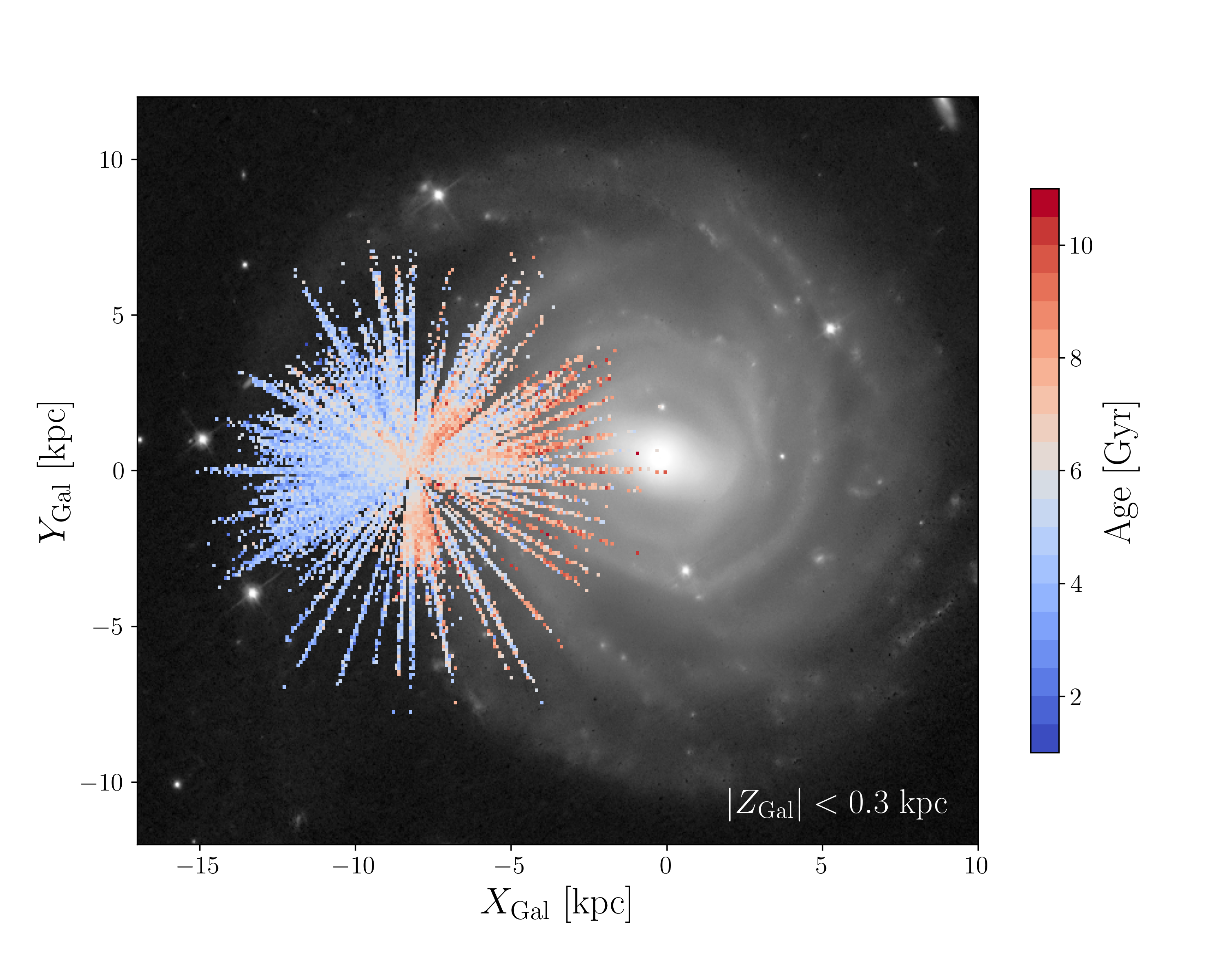}
\caption{\label{fig:xy_age2} 
Median age as a function of position in Galactocentric Cartesian coordinates $\{X_{\rm Gal}, Y_{\rm Gal}\}$ (top-down view of the Galaxy), showing a slice of Galactic disc close to the Galactic plane ($|Z_{\rm Gal}|<0.3$ kpc). An $r$-band image of the nearby face-on spiral galaxy UGC 12158 is shown in the background for illustration. 
} 
\end{center}
\end{figure*}

For illustration purposes, we include two additional figures highlighting the Galactic coverage of the present APOGEE spectroscopic age data. Figure \ref{fig:rz_age2} shows the median age per pixel in cylindrical coordinates, similar to Fig. \ref{fig:rz_age}, but now using equally-scaled axes, and with a Milky-Way-like edge-on galaxy (NGC 891) as a background image. Figure \ref{fig:xy_age2} shows a top-down view of the Galactic disc colour-coded by median age, with the face-on Milky-Way analogue UGC 12158 in the background.

\section{Additional validation} \label{sec:appendix}

\begin{figure}
\begin{center} 
\includegraphics[width=.49\textwidth]{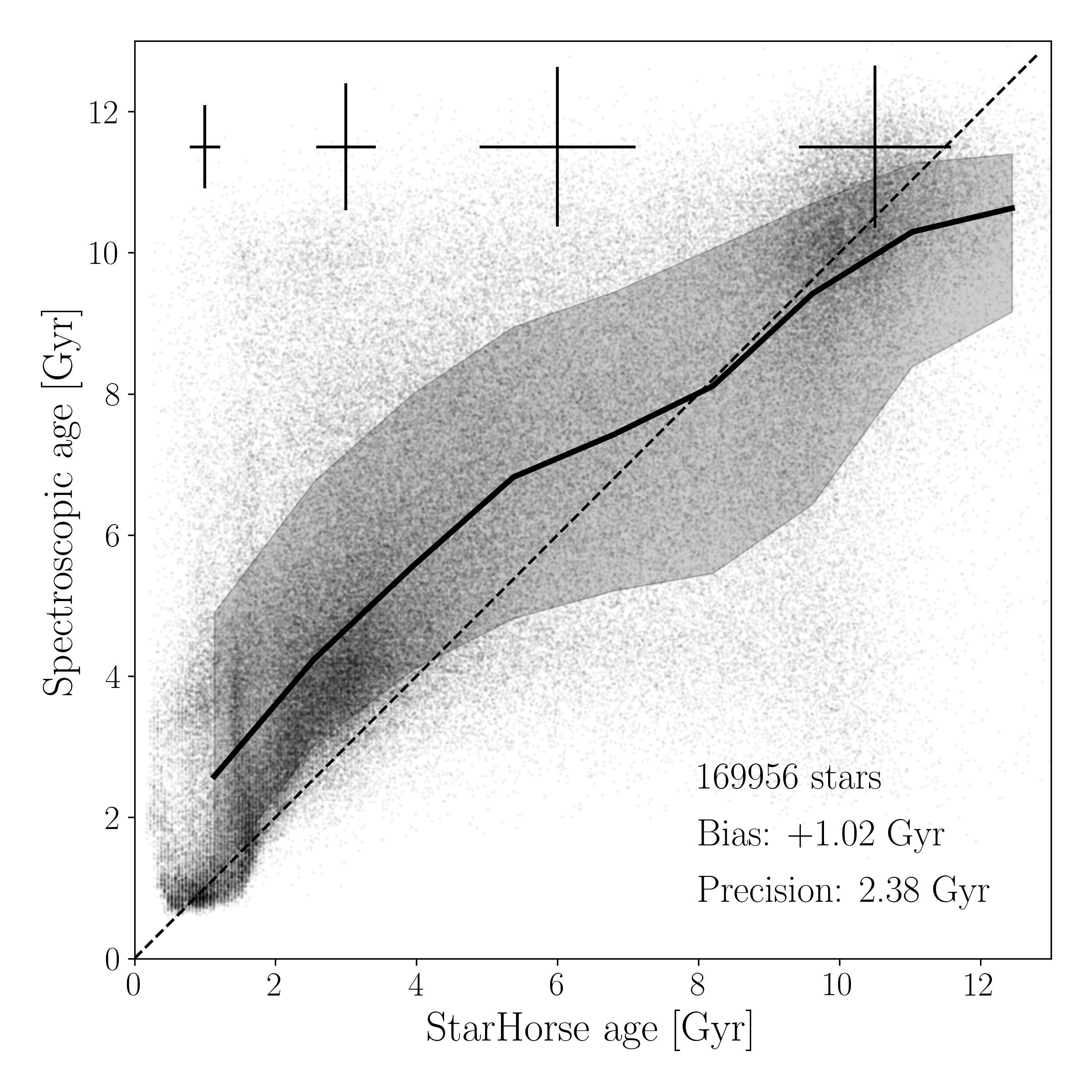}
\includegraphics[width=.49\textwidth]{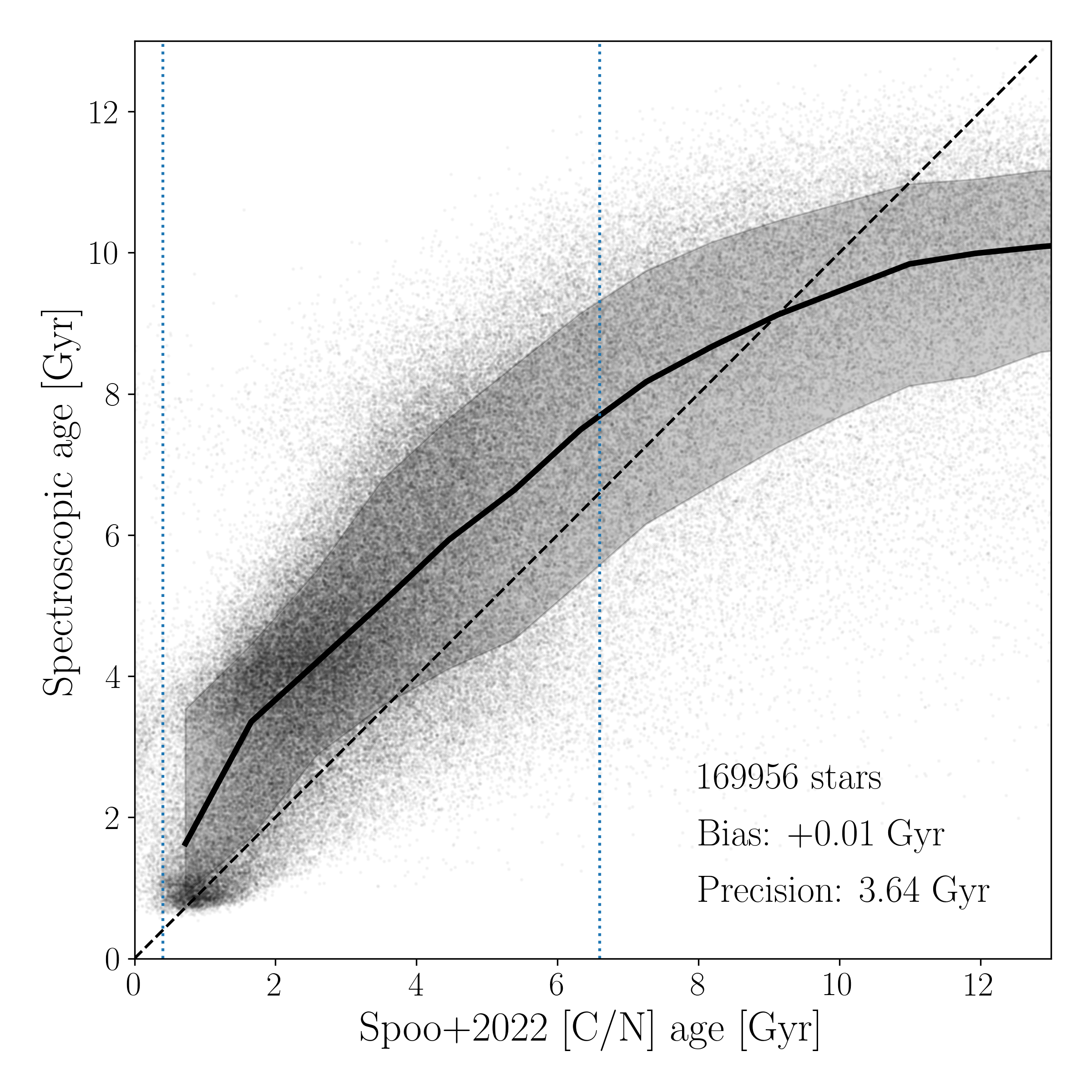}\\
\caption{\label{fig:age_comp_2} 
Additional comparisons of our uncalibrated age estimates with other field-star age estimates for APOGEE DR17. Top: Unpublished red-giant star age estimates from {\tt StarHorse} \citep{Queiroz2023}. Bottom: Empirically calibrated DR17 [C/N] age estimates from \citet{Spoo2022}. The dotted vertical lines enclose the age regime in which their [C/N] age calibration is valid.
}
\end{center}
\end{figure}

\subsection{{\tt StarHorse} ages for APOGEE DR17\citep{Queiroz2023}}\label{sec:sh}

The top panel of Fig. \ref{fig:age_comp_2} shows a direct comparison with the age estimates of \citet{Queiroz2023} for the APOGEE DR17 catalogue. The authors used the Bayesian isochrone-fitting code {\tt StarHorse} \citep{Queiroz2018, Queiroz2020} to simultaneously obtain stellar parameters for the APOGEE DR17 catalogue, cross-matched with {\it Gaia} EDR3 astrometry \citep{GaiaCollaboration2021} as well as several photometric catalogues. Qualitatively, the comparison reveals a similar systematic trend as the comparison to the seismic and open-cluster ages shown in the top panels of Fig. \ref{fig:age_comp_5}: good concordance for the youngest ages followed by prominent positive ($\sim +1$ Gyr) residuals for ages between 2 and 4 Gyr, after which the trend regresses towards zero residuals again. We note, however, that \citet{Queiroz2023} only published the ages inferred for sub-giant and main-sequence turn-off stars, since isochrone ages for red giants are still prone to sizeable systematic uncertainties -- which also explains the larger scatter. 

\subsection{[C/N] ages \citep{Spoo2022}}\label{sec:spoo}

Since the APOGEE discovery of a systematic trend of [C/N] with the position in the [Mg/Fe] vs. [Fe/H] diagram and, therefore, possibly stellar age \citep{Masseron2015}, several authors have tried to calibrate a relation between the observed [C/N] ratio and the age of red giant stars (e.g. \citealt{Martig2016, Ness2016, Casali2019}). To obtain a solid calibration for evolved stars, \citet{Spoo2022} used the APOGEE DR17 data from the Open Cluster Chemical Abundances and Mapping survey (OCCAM; \citealt{Frinchaboy2013, Myers2022}). Using 94 open clusters tagged as "highly reliable", they obtained the following relationship:
$$\mathrm{log}{[\mathrm{Age}(\mathrm{yr})]}_{\mathrm{DR}17}=10.14\,(\pm 0.08)+2.23(\pm 0.19)\cdot[{\rm{C}}/{\rm{N}}],$$
valid for ages between 0.4 Gyr and 6.6 Gyr. 

The lower panel of Fig. \ref{fig:age_comp_2} shows the comparison of our spectroscopic age estimates with the calibration obtained by \citet{Spoo2022}. Again, we find similar trends as in those visible in Fig. \ref{fig:age_comp_5}, albeit with much larger scatter, which is expected, since the empirical relation of \citet{Spoo2022} only uses two abundance measurements to infer an age estimate. 

\subsection{Age-abundance relations}\label{sec:age-mgfe}

\begin{figure}
\begin{center} 
\includegraphics[width=.495\textwidth]{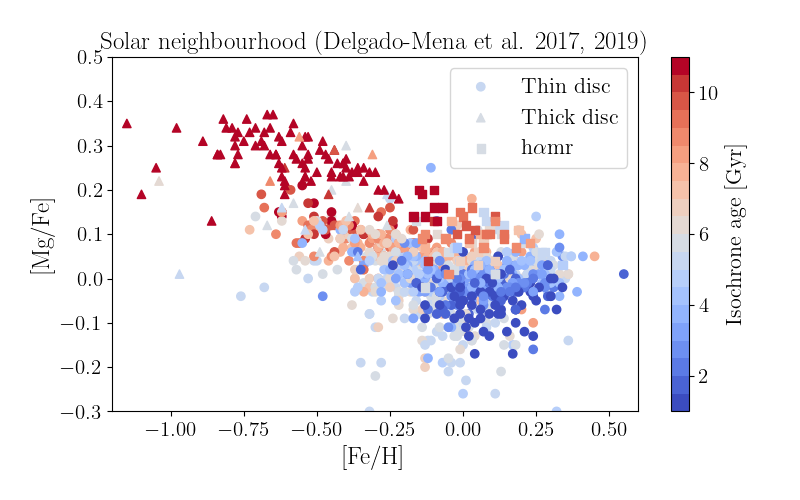}\\
\includegraphics[width=.495\textwidth]{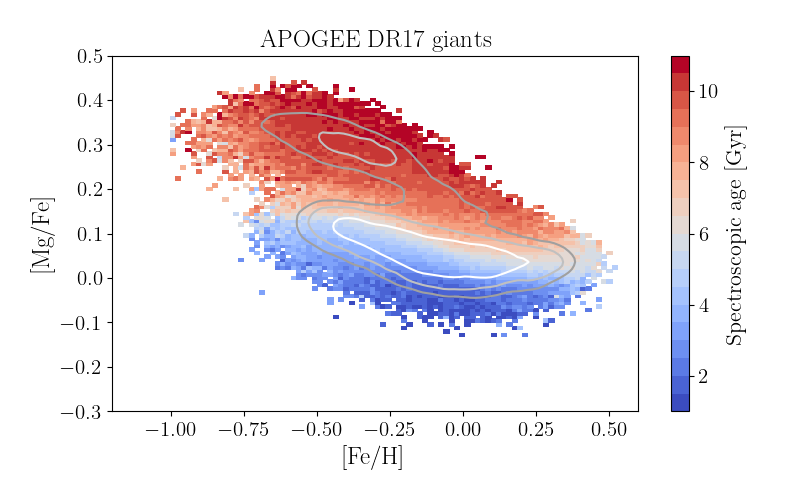}\\
\caption{\label{fig:mgfe_age} 
[Mg/Fe] vs. [Fe/H] Tinsley diagram colour-coded by stellar age. Top panel: scatter plot for the high-resolution solar-vicinity sample of \citet{DelgadoMena2019}, using the same symbols as in the original publication. Bottom panel: Median age per pixel in the [Mg/Fe] vs. [Fe/H] diagram for our 200\,000 APOGEE DR17 giants. Overplotted are iso-density contours corresponding to 40, 80, and 160 stars per bin. } 
\end{center}
\end{figure}

\begin{figure}
\begin{center} 
\includegraphics[width=.495\textwidth]{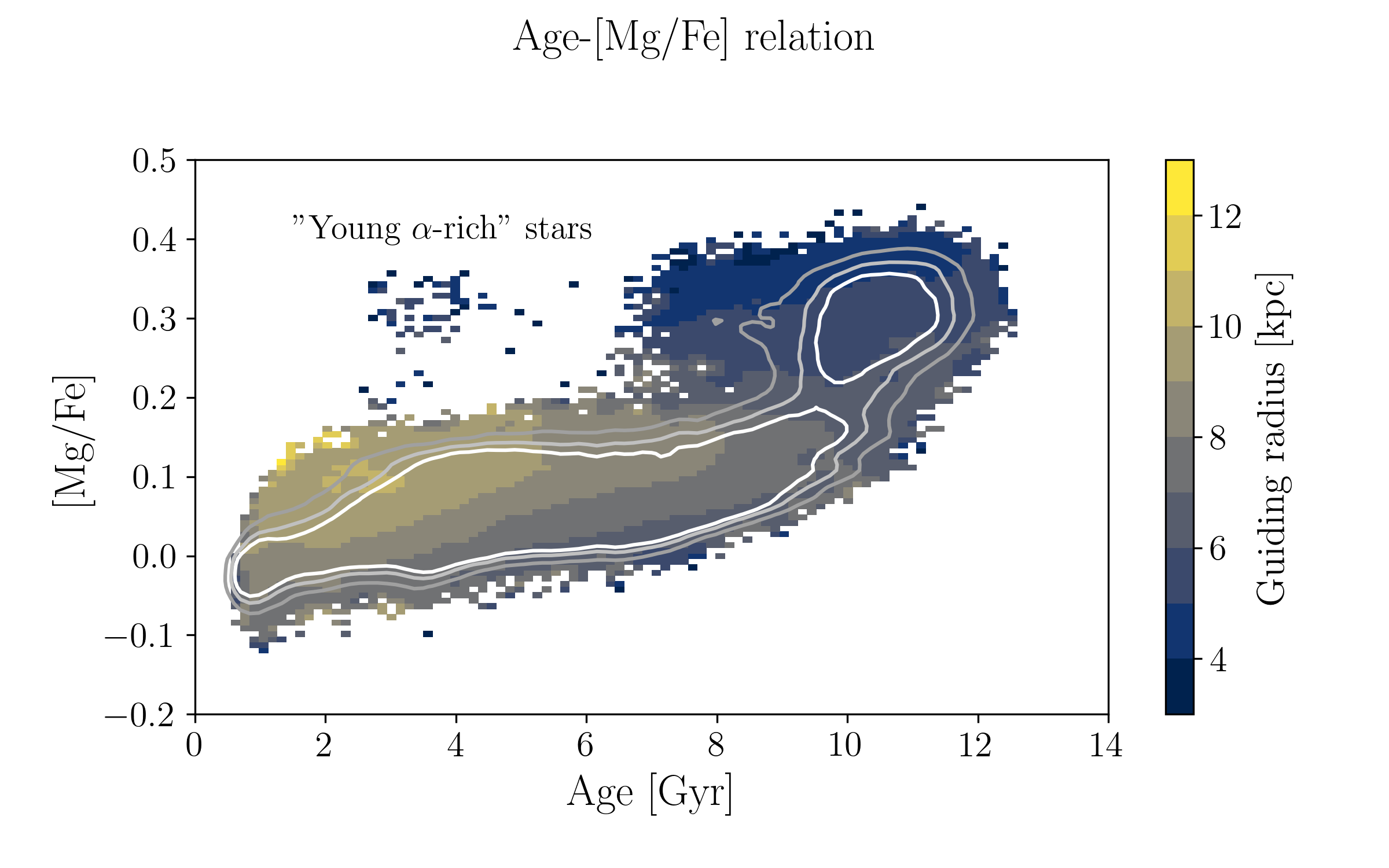}\\
\caption{\label{fig:agemg_R} 
Age-[Mg/Fe] relation, in the same style as Fig. \ref{fig:agefe_R}. The small population of young $\alpha$-rich stars is highlighted.} 
\end{center}
\end{figure}


From Fig. \ref{fig:shap} we learned that the five most influential features for our spectroscopic age estimator are [C/Fe], $T_{\rm eff}$, [N/Fe], [Mg/Fe], and $\log g$. 
Therefore, before we discuss some of the abundance trends with age, a note of caution is due: ideally one wishes to study stellar abundance ratios as a function of an age estimate that is both accurate and independent from the spectroscopically inferred abundances. In our case, however, the inferred ages rely on the measured abundances, and therefore our derived abundance ratio (e.g. [Mg/Fe], Fig. \ref{fig:agemg_R}) vs. age trends should only be viewed as sanity checks, not as proofs of the validity of the concept of chemical clocks.

Figure \ref{fig:mgfe_age} shows the [Mg/Fe] vs. [Fe/H] diagram colour-coded by age, for the solar-vicinity sample of \citet{DelgadoMena2017, DelgadoMena2019} and for our APOGEE DR17 giant sample. The figure demonstrates that our data reproduce the expected age trends in this diagram, and that these trends persist over the large portion of the Galactic disc covered by our sample. It also confirms that the so-called high-$\alpha$ metal-rich population (h$\alpha$mr, \citealt{Adibekyan2011}) is dominantly old (between 9 and 11 Gyr), indicating that this population was most likely formed in the very inner disc and/or bulge and migrated outwards \citep[e.g.][]{Anders2018, Queiroz2021, Ratcliffe2023}.

As a typical example of an elemental-abundance ratio that is often referred to as a chemical clock, we show the age-[Mg/Fe] relation for the full APOGEE sample in Fig. \ref{fig:agemg_R}. The iso-density contours in this plot show that, in accordance with expectations from classical chemical evolution models \citep[e.g.][]{Chiappini2009}, the relationship is almost flat, except for the oldest ages, where we see a sharp transition from the high-[Mg/Fe] regime to the low-[Mg/Fe] regime, related to the onset of type-Ia supernovae \citep[e.g.][]{Mannucci2006, Mennekens2010}. The colour code in Fig. \ref{fig:agemg_R} shows how the age-[Mg/Fe] varies with guiding radius, from a shorter and steeper relation in the outer parts of the disc (yellow colours) to a very flat relation (for ages $< 8$ Gyr) inside the solar circle. 

Figure \ref{fig:agemg_R} also highlights the presence of a small population of so-called young $\alpha$-rich stars. These stars, first discovered in APOGEE data \citep{Chiappini2015, Martig2015}, have since been demonstrated to be products of binary stellar evolution -- they have been rejuvenated by mass transfer from a stellar companion \citep[e.g.][]{Jofre2016, Yong2016, Fuhrmann2017, Fuhrmann2018, Hekker2019, Sun2020, Zhang2021, Jofre2022}. Therefore, their (isochrone-derived) ages are catastrophically underestimated. 
The presence of the few young $\alpha$-rich stars in the full APOGEE sample (not present in the training set) reminds us that our age estimates are not completely based on the stellar photospheric compositions alone, but also depend on effective temperature and gravity, which may also mimic a younger stellar age for a given chemical composition. We therefore flag these stars as problematic in our catalogue (see Sect. \ref{sec:datamodel}).

\section{Fitting the radial [Fe/H] distributions}\label{sec:appendix2}

In Figs. \ref{fig:dfedr_age} and \ref{fig:sigfe_age} (Sect. \ref{sec:grad}) we presented the results of Bayesian fits to the radial [Fe/H] abundance distributions in age bins of 0.5 Gyr. The method (in particular, the implementation of the likelihood and posterior) is described in \citet{Anders2017a}. In this appendix we show the detailed results of these fits for some age bins. 

\begin{figure*}
\begin{center} 
\includegraphics[width=.55\textwidth]{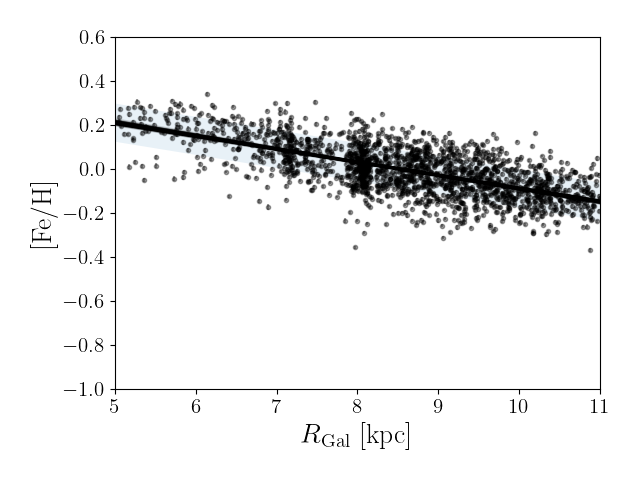}
\includegraphics[width=.4\textwidth]{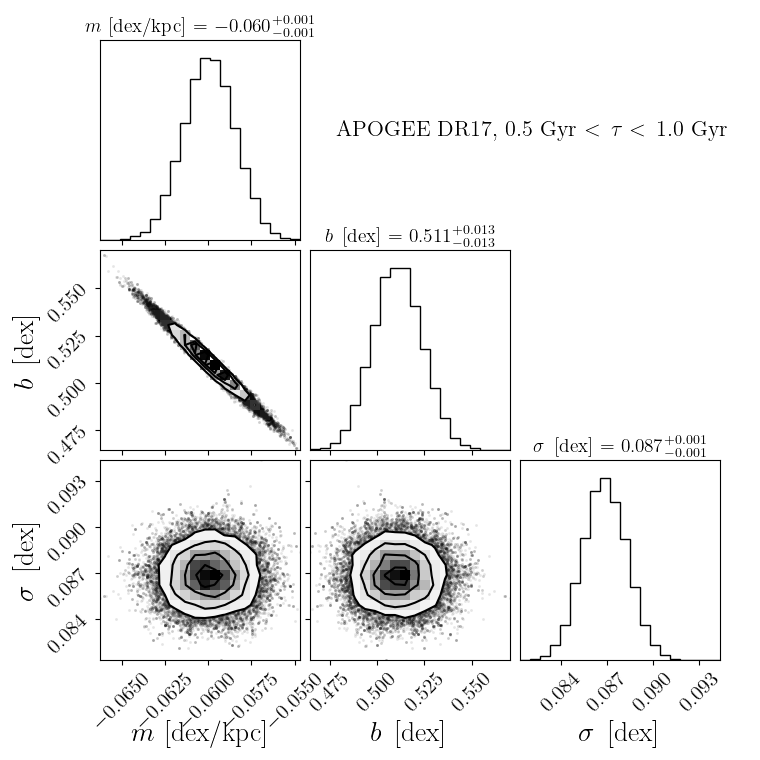}
\includegraphics[width=.55\textwidth]{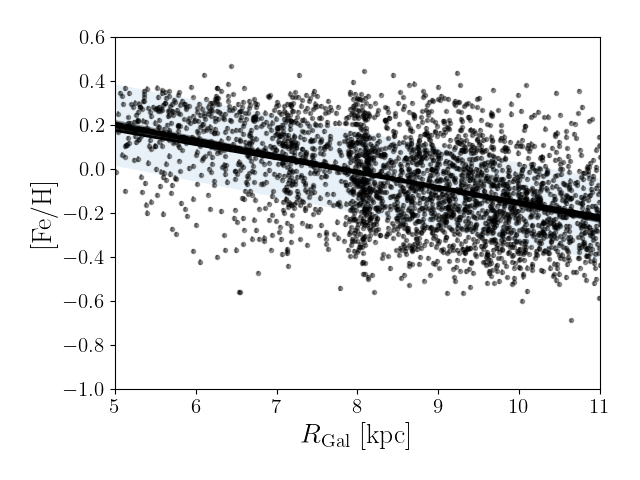}
\includegraphics[width=.4\textwidth]{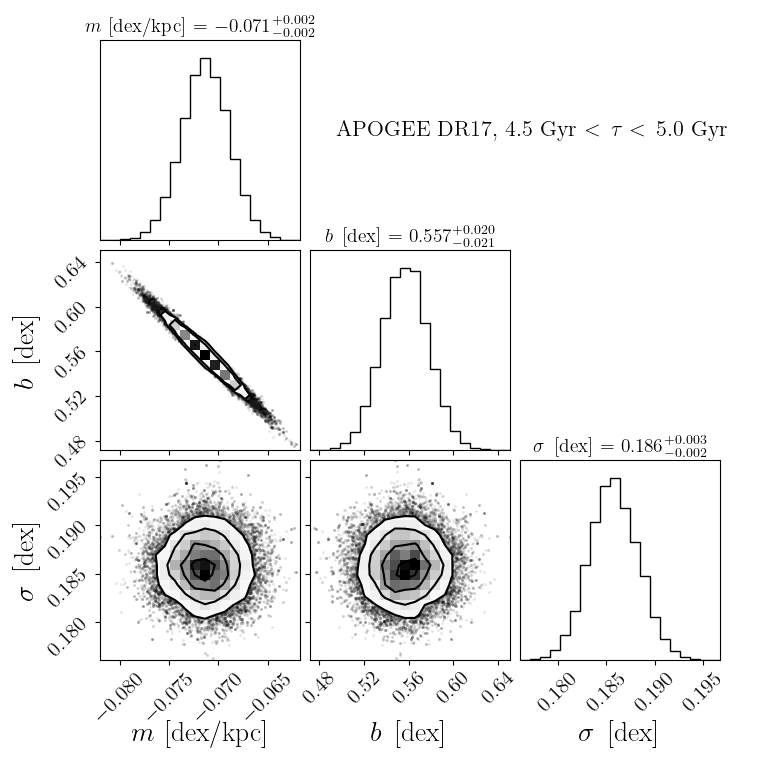}
\includegraphics[width=.55\textwidth]{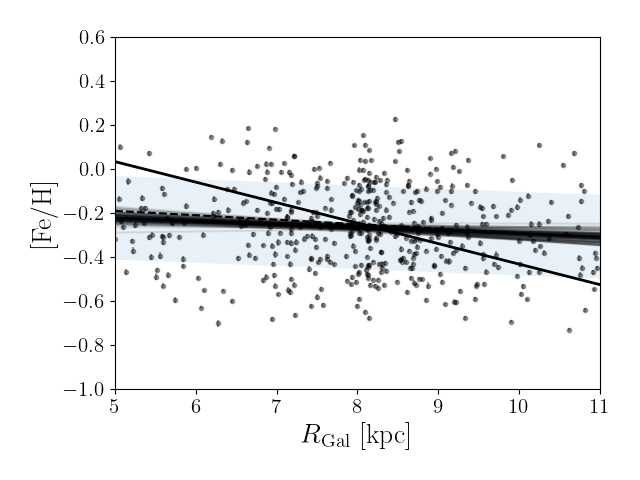}
\includegraphics[width=.4\textwidth]{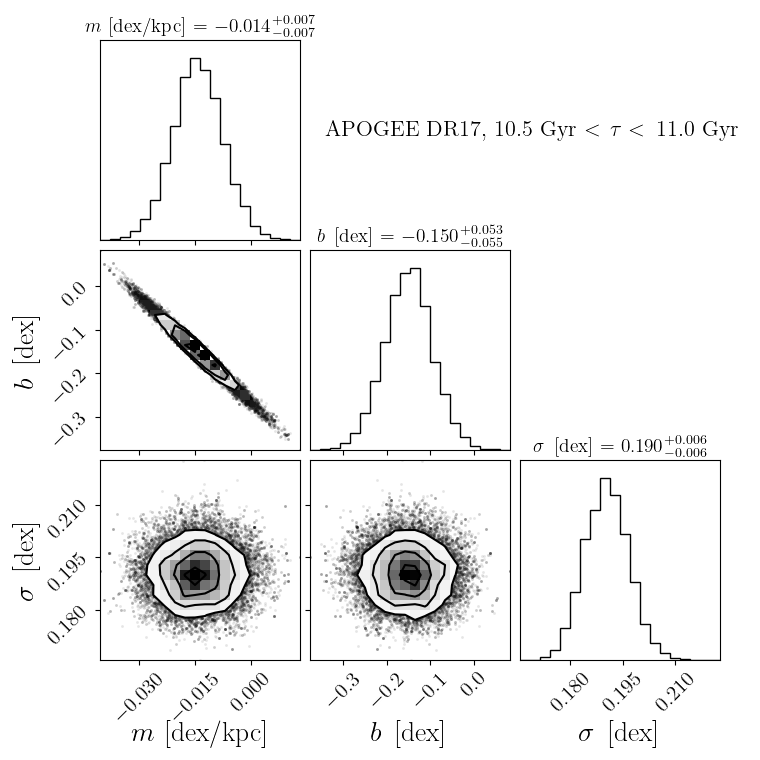}
\caption{\label{fig:agebins_gradfit} 
Examples of the fits to the [Fe/H] vs. $R_{\rm Gal}$ distributions for three age bins (from top to bottom: 0.5-1 Gyr, 4.5-5 Gyr, 10.5-11 Gyr). Left panels: distribution of the red-giant stars in each age bin. The thick black line shows the result of a naïve least-squares linear fit. The thin grey lines show 30 realisations drawn from the linear gradient + intrinsic scatter posterior, while the shaded band corresponds the $1\sigma$ dispersion around the gradient. Right panels: posterior distributions of the fit parameters ($m=\partial$[Fe/H]$/\partial R$, $b$ (intercept at $R=0$), and $\sigma$ (intrinsic [Fe/H] dispersion).} 
\end{center}
\end{figure*}

\end{appendix}

\end{document}